\documentclass[preprintnumbers, floatfix, preprintnumbers, letterpaper, superscriptaddress,nofootinbib]{revtex4}
\pdfoutput=1
\usepackage{graphicx}
\usepackage{microtype}
\usepackage{amsmath}
\usepackage{amssymb}
\usepackage{subfigure}
\usepackage{hyperref}
\usepackage{url}
\usepackage{xcolor}
\usepackage{color}
\usepackage{mathrsfs}
\usepackage{calrsfs}
\usepackage{amsfonts}
\usepackage{latexsym}
\usepackage{ragged2e}
\usepackage{epstopdf}
\usepackage{textcomp}
\usepackage{phaistos}
\usepackage{ulem}

\makeatletter
\renewcommand\@makefnmark{\hbox{\@textsuperscript{\normalfont\color{purple}\@thefnmark}}}
\renewcommand\@makefntext[1]{%
  \parindent 1em\noindent
            \hb@xt@1.8em{%
                \hss\@textsuperscript{\normalfont\@thefnmark}}#1}
\makeatother

\usepackage{caption}
\DeclareCaptionJustification{justified}{\leftskip=0pt \rightskip=0pt \parfillskip=0pt plus 1fil}
\definecolor{vividviolet}{rgb}{0.62, 0.0, 1.0}
\definecolor{amaranth}{rgb}{0.9, 0.17, 0.31}
\definecolor{palatinateblue}{rgb}{0.15, 0.23, 0.89}
\definecolor{brightpink}{rgb}{1.0, 0.0, 0.5}
\definecolor{cornflowerblue}{rgb}{0.39, 0.58, 0.93}
\definecolor{deepcarminepink}{rgb}{0.94, 0.19, 0.22}
\definecolor{radicalred}{rgb}{1.0, 0.21, 0.37}

\hypersetup{ linktoc=all,
    colorlinks, linkcolor={palatinateblue},
    citecolor={brightpink}, urlcolor={amaranth}
}

\graphicspath{{Images/}}

\def\sideremark#1{\ifvmode\leavevmode\fi\vadjust{\vbox to0pt{\vss
 \hbox to 0pt{\hskip\hsize\hskip1em
 \vbox{\hsize1.5cm\tiny\raggedright\pretolerance10000
 \noindent #1\hfill}\hss}\vbox to8pt{\vfil}\vss}}}%
\begin{document}

\title{Properties of Scalar Hairy Black Holes and Scalarons with Asymmetric Potential}

\author{Xiao Yan \surname{Chew}}
\email{xychew998@gmail.com}
\affiliation{Department of Physics Education, Pusan National University, Busan 46241, Republic of Korea}
\affiliation{Research Center for Dielectric and Advanced Matter Physics, Pusan National University, Busan 46241, Republic of Korea}

\author{Dong-han \surname{Yeom}}
\email{innocent.yeom@gmail.com}
\affiliation{Department of Physics Education, Pusan National University, Busan 46241, Republic of Korea}
\affiliation{Research Center for Dielectric and Advanced Matter Physics, Pusan National University, Busan 46241, Republic of Korea}

\author{Jose Luis \surname{Bl\'azquez-Salcedo}}
\email{jlblaz01@ucm.es}
\affiliation{Departamento de F\'sica Te\'orica and IPARCOS, Universidad Complutense de Madrid, E-28040 Madrid, Spain}

\begin{abstract}
In this paper we study the properties of black holes and scalarons in Einstein gravity when it is minimally coupled to a scalar field $\phi$ with an asymmetric potential $V(\phi)$, constructed in [Phys. Rev. D \textbf{73} (2006), 084002] a few decades ago. $V(\phi)$ has been applied in the cosmology to describe the quantum tunneling process from the false vacuum to the true vacuum and contains a local maximum, a local minimum (false vacuum) and a global minimum (true vacuum). In particular we focus on the asymptotically flat solutions, which can be constructed by fixing appropriately the local minimum of $V$.
A branch of hairy black holes solutions emerge from the Schwarzschild black hole, and we study the domain of existence of such configurations. They can reach to a particle-like solution in the small horizon limit, i.e. the scalarons.
We study the stability of black holes and scalarons, showing that all of them are unstable under radial perturbations. 
\end{abstract}

\maketitle

\section{Introduction}

The ``No hair" theorem states that the properties of black holes are only described by the mass, angular momentum and electrical charge after gravitational collapse or any dynamical perturbations of black holes, since they approach the stationary limit. However, the ``No hair" theorem can be circumvented 
under the right conditions. For example,
the existence of particle-like solution for SU(2) Einstein-Yang-Mills theory shown by Bartnik and Mckinnon \cite{Bartnik:1988am} had led to the construction of non-Abelian hairy black holes \cite{Bizon:1990sr,Volkov:1990sva,Kuenzle:1990is,Greene:1992fw,Lavrelashvili:1992ia} which don't obey the ``No hair theorem" anymore. 
One of 
the simplest way to circumvent the ``No hair" theorem is to minimally couple Einstein gravity 
with a scalar field, introducing a scalar potential which is not strictly positive such that the weak energy condition is violated \cite{Herdeiro:2015waa}.
In \cite{Corichi:2005pa} spherically symmetric and asymptotically flat hairy black holes were constructed by employing a scalar potential which has a global minimum, a local minimum and a local maximum (asymmetric potential). 
They obtained the asymptotically flat black holes by fixing the local minimum of potential to zero, to mainly study the empirical mass formula of such black holes \cite{Corichi:2005pa} and later also generalize their model to non-minimally coupled scalar field with gravity \cite{Nucamendi:1995ex}. However, the properties of black holes such as the Hawking temperature and mass had not been investigated systematically 
in terms of the parameters of the scalar potential.
In this work we investigate the properties of these solutions by fixing the global minimum of the potential and varying the local maximum. This allows us to generate a branch of hairy black holes that bifurcate from the Schwarzschild black holes. 
See \cite{Bechmann:1995sa,Dennhardt:1996cz,Bronnikov:2001ah,Martinez:2004nb,Winstanley:2005fu,Nikonov:2008zz,Anabalon:2012ih,Gao:2021ubl,Karakasis:2021rpn} for similar constructions of scalar hairy black holes.

When there is a scalar field with an asymmetric potential, a quantum tunneling from a false vacuum (a local minimum) to a true vacuum (a global minimum) is allowed. With an $O(4)$-symmetric metric ansatz, the Coleman-De Luccia instantons explain such a tunneling process via nucleation of a bubble \cite{Coleman:1980aw}. Even we extend to the spherical symmetry, again, one can build a solution that explains a tunneling process \cite{Masoumi:2012yy}. For both cases, inside is a true-vacuum region, and after the nucleation, the bubble should expand over the spacetime; otherwise, the scalar field combination is, in general, unstable. Such a bubble may explain the phase transition of the early universe cosmology \cite{DeLuca:2021mlh,Perivolaropoulos:2022txg,Braden:2022odm,Ekstedt:2022ceo,Cruz:2022ext,Vicentini:2022pra}; 
also, some interactions between bubbles may be a source of gravitational waves \cite{Kosowsky:1991ua,Kim:2014ara,Lee:2021nwg}.

On the other hand, one can also say that the same solution can be interpreted as a kind of (unstable) scalarons or stationary hairy black hole solutions. In order to provide a smooth field configuration at the horizon over the Euclidean manifold, one needs to provide the Dirichlet boundary condition at the event horizon. If we generalize this boundary condition and focus only on the astrophysical aspects, it is still allowed to provide more generic boundary conditions than the pure Dirichlet boundary condition \cite{Masoumi:2012yy,Gregory:2020cvy}. This is the case that we will investigate in the present paper.

This paper is organized as follows. In Sec.~\ref{sec:th}, we briefly introduce our theoretical setup comprising the Lagrangian and the metric ansatz. Then, we derive the set of coupled differential equations and study the asymptotic behavior of the functions. In Sec.~\ref{sec:prop}, we briefly introduce the quantities of interest for the black holes. In Sec.~\ref{sec:lin}, we study the stability of the hairy black holes and scalarons by calculating the unstable mode of the radial perturbations of the metric and the scalar field. %
In Sec~\ref{sec:res}, we present and discuss our numerical findings. Finally, in Sec.~\ref{sec:con}, we summarize our work and present an outlook.

\section{Theoretical Setting} \label{sec:th}

\subsection{Theory and Ans\"atze}

\begin{figure}[t!]
\centering
\mbox{
\includegraphics[trim=10mm 170mm 10mm 15mm,scale=0.45]{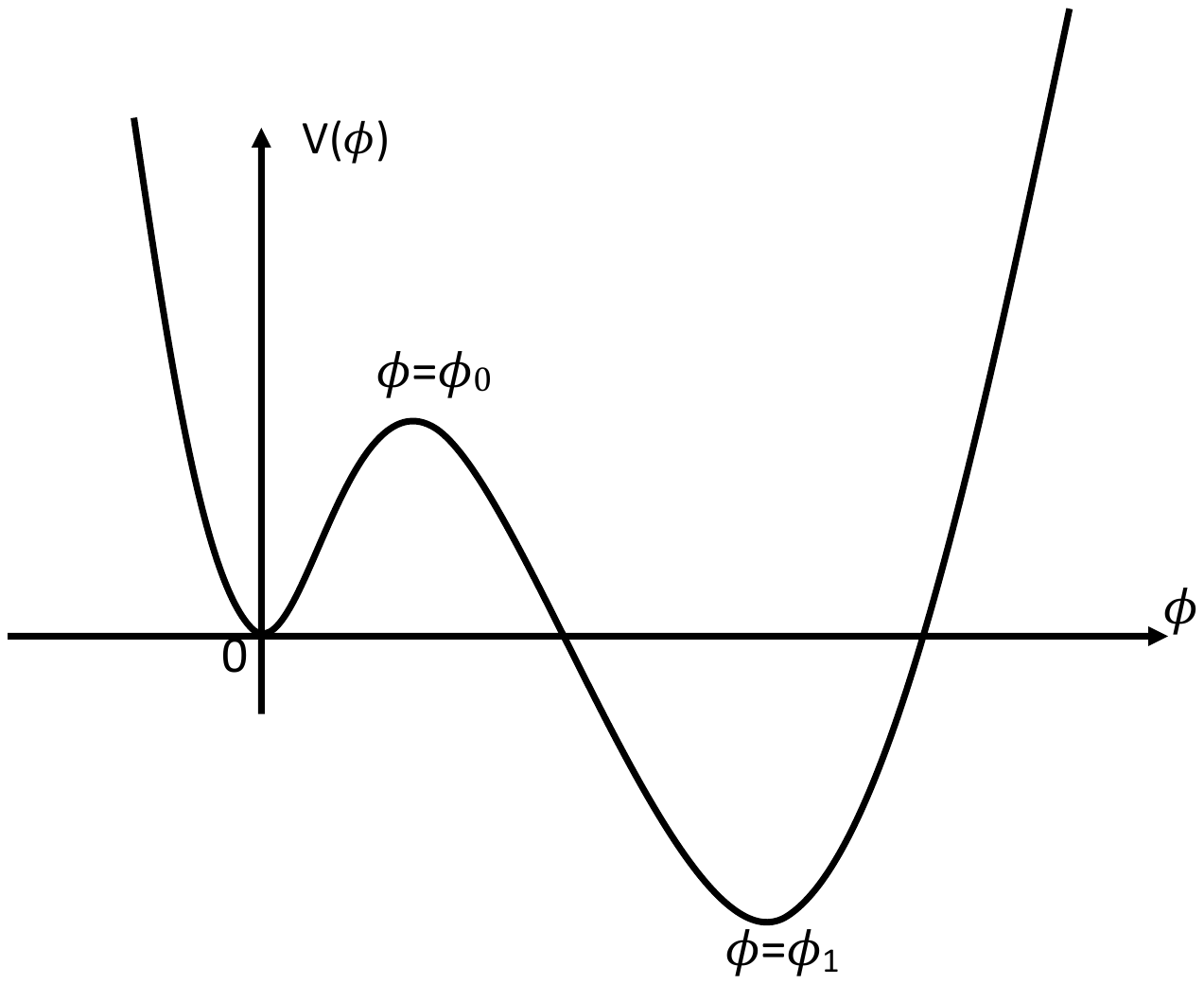}
}
\caption{An illustration of a generic asymmetric scalar potential $V(\phi)$.}
\label{plot_Vphi}
\end{figure}

The action for Einstein gravity minimally couples with an asymmetric potential $V(\phi)$ of a scalar field $\phi$ is given by \cite{Corichi:2005pa}
\begin{equation} \label{EHaction}
 S=  \int d^4 x \sqrt{-g}  \left[  \frac{R}{16 \pi G} - \frac{1}{2} \nabla_\mu \phi \nabla^\mu \phi - V(\phi) \right]  \,,
\end{equation}
where
\begin{equation}
 V(\phi) = \frac{V_0}{12} \left( \phi - a \right)^2 \left[ 3 \left( \phi-a\right)^2 - 4 (\phi-a) (\phi_0 + \phi_1) + 6 \phi_0 \phi_1  \right] \,,
\end{equation}
with $a$, $V_0$, $\phi_0$ and $\phi_1$ are the constants. As shown in Fig \ref{plot_Vphi}, the appearance of cubic term $\phi^3$ causes the potential to take the asymmetric shape. If the cubic term in the potential disappears, then the potential is Higgs like with two degenerate minima. Here the constant $a$ is the local minimum of potential, $\phi_0$ is the local maximum of potential and $\phi_1$ is the global minimum of potential. Note that $0 < 2 \phi_0 < \phi_1$. In cosmology, this potential can be used to explain a quantum tunneling process
from the false vacuum $a$ to the true vacuum $\phi_1$. In this paper, we choose $a=0$ such that the hairy black holes is asymptotically flat at the spatial infinity. However, one could obtain a Schwarzschild-AdS like solutions if $a$ is non-zero \cite{Corichi:2005pa}. 

The variation of the action with respect to the metric and scalar field yields the Einstein equation and Klein-Gordon (KG) equation, respectively
\begin{equation} 
 R_{\mu \nu} - \frac{1}{2} g_{\mu \nu} R = 8 \pi G T_{\mu \nu} \,,  \quad  \nabla_\mu \nabla^\mu \phi  =  \frac{d V(\phi)}{d \phi} \,,  
\end{equation}
where the stress-energy tensor $T_{\mu \nu}$ is given by
\begin{equation}
 T_{\mu \nu} =  -g_{\mu \nu} \left(  \frac{1}{2} \nabla_\alpha \phi \nabla^\alpha \phi + V(\phi) \right) + \nabla_\mu \phi \nabla_\nu \phi \,.
\end{equation}

We employ the following spherically symmetric Ansatz to construct the particle-like and black holes solutions,
\begin{equation}  \label{line_element}
ds^2 = - N(r) e^{-2 \sigma(r)} dt^2 + \frac{dr^2}{N(r)} + r^2  \left( d \theta^2+\sin^2 \theta d\varphi^2 \right) \,, 
\end{equation}
where $N(r)=1-2 m(r)/r$ with $m(r)$ is the Misner-Sharp mass function. Note that $m(\infty)=M$, the total mass of the configuration.

\subsection{Justification on the Existence of Hairy Black Holes}

\begin{figure}[t!]
\centering
 \includegraphics[angle =0,scale=0.8]{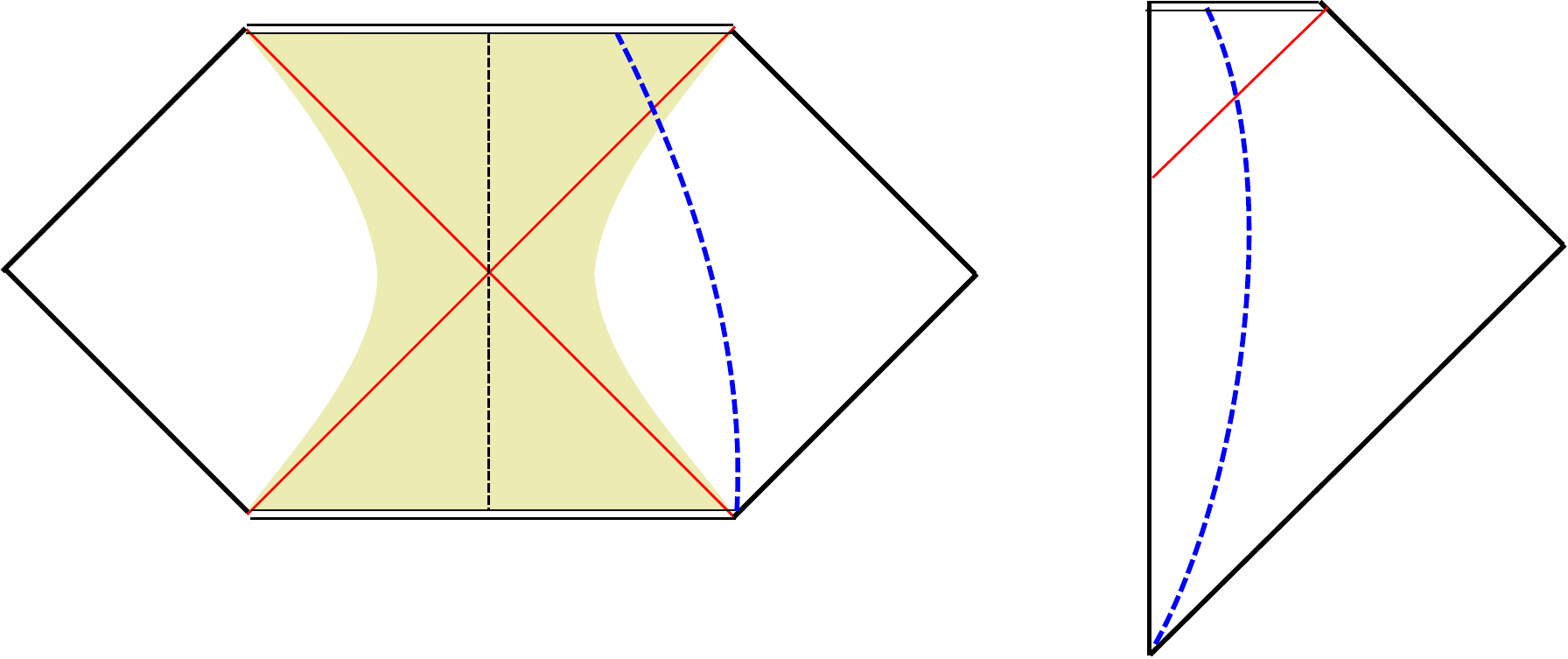}
\caption{Left: The maximally extended causal structure for hairy black holes. The gray-colored part denotes a region for negative vacuum energy. The black dashed curve denotes a cusp surface of the scalar field and the blue dashed curve denotes a star surface. Right: The physically sensible interpretation of the hairy black hole does not include any cusp region.
 }
\label{plot_causal}
\end{figure}

Here we briefly justify the existence of hairy black holes by referring to \cite{Herdeiro:2015waa}. We multiply $\phi$ to the Klein-Gordon equation and integrate it from the black hole horizon to infinity,
\begin{equation}
 \int_{r_H}^{\infty} d^4 x \sqrt{-g} \left[ \phi  \nabla_\mu\nabla^\mu \phi - \phi  \frac{d V(\phi)}{d \phi}   \right]  = 0 \,.
\end{equation}

We integrate the first term in above expression by parts and obtain,
\begin{equation}
 \int_{r_H}^{\infty} d^4 x \sqrt{-g} \left[ - \nabla_\mu \phi \nabla^\mu \phi - \phi  \frac{d V(\phi)}{d \phi}   \right] + \int_{\mathcal{H}} d^3\sigma n^\mu \phi \nabla_\mu \phi  = 0 \,,
\end{equation}
where $n^\mu$ is the normal vector on the Killing horizon. The second term in the above expression is the boundary term which vanishes when we apply the boundary conditions for the scalar field at the horizon with $n^\mu\nabla_\mu\phi=0$ and demand the scalar field falls off at infinity. Hence, we are left with
\begin{equation}
 \int_{r_H}^{\infty} d^4 x \sqrt{-g} \left[  \nabla_\mu \phi \nabla^\mu \phi + \phi  \frac{d V}{d \phi}   \right] =0 \,,
\end{equation}
Here $ \nabla_\mu \phi \nabla^\mu \phi \geq 0$ because the gradient of $\phi$ is perpendicular to both Killing vectors and thus has to be spacelike or zero. Then in order to obtain a regular and nontrivial hairy black hole, the term $\phi  \frac{d V}{d \phi}  \leq 0$. In our case, we choose $\phi$ to be always greater than zero, while the potential $V(\phi)$ is negative in some region; therefore we allow for the existence of nontrivial scalar hairy black holes. 

Furthermore, one can multiply the Klein-Gordon equation by $\frac{d V}{d \phi}$ and repeat again the above procedure, obtaining
\begin{equation}
 \int_{r_H}^{\infty} d^4 x \sqrt{-g} \left[  \frac{d^2 V}{d \phi^2} \nabla_\mu \phi \nabla^\mu \phi  + \left(   \frac{d V}{d \phi}  \right)^2    \right] = 0   \,.
\end{equation}
In order to make the terms in the square bracket to vanish non-trivially, it is clear that $\frac{d^2 V}{d \phi^2} < 0 $, a condition that is also satisfied in our case. Note that in this derivation, it is not necessary to use the Einstein equation.

On the other hand, we can also see that the weak energy condition can be violated since $V$ possesses a global minimum with $V < 0$ in some regions of $\phi$, 
\begin{equation}
 \rho = - T^t_{t} = \frac{N}{2} \phi'^2 + V  \,.
\end{equation}
The violation of weak energy condition leads to the violation of the strong energy condition (the opposite not being necessarily true). Moreover, we could also use the Virial identity to reach the same conclusion, that is, $V \leq 0$ in some region, but for this analysis it is necessary to introduce the metric Ansatz into the action.

\begin{figure}[t!]
\centering
\mbox{
(a)
 \includegraphics[angle =-90,scale=0.33]{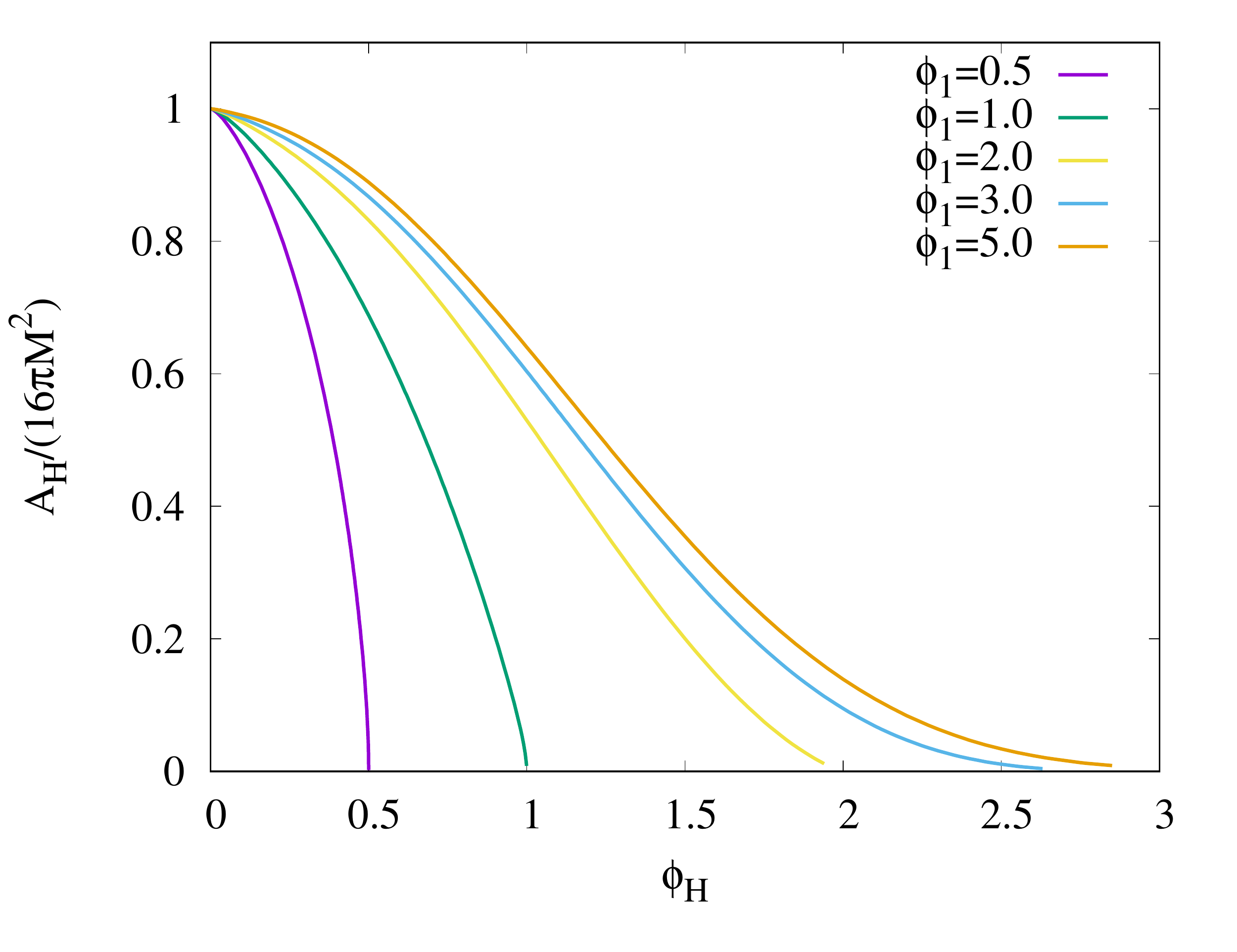}
(b)
 \includegraphics[angle =-90,scale=0.33]{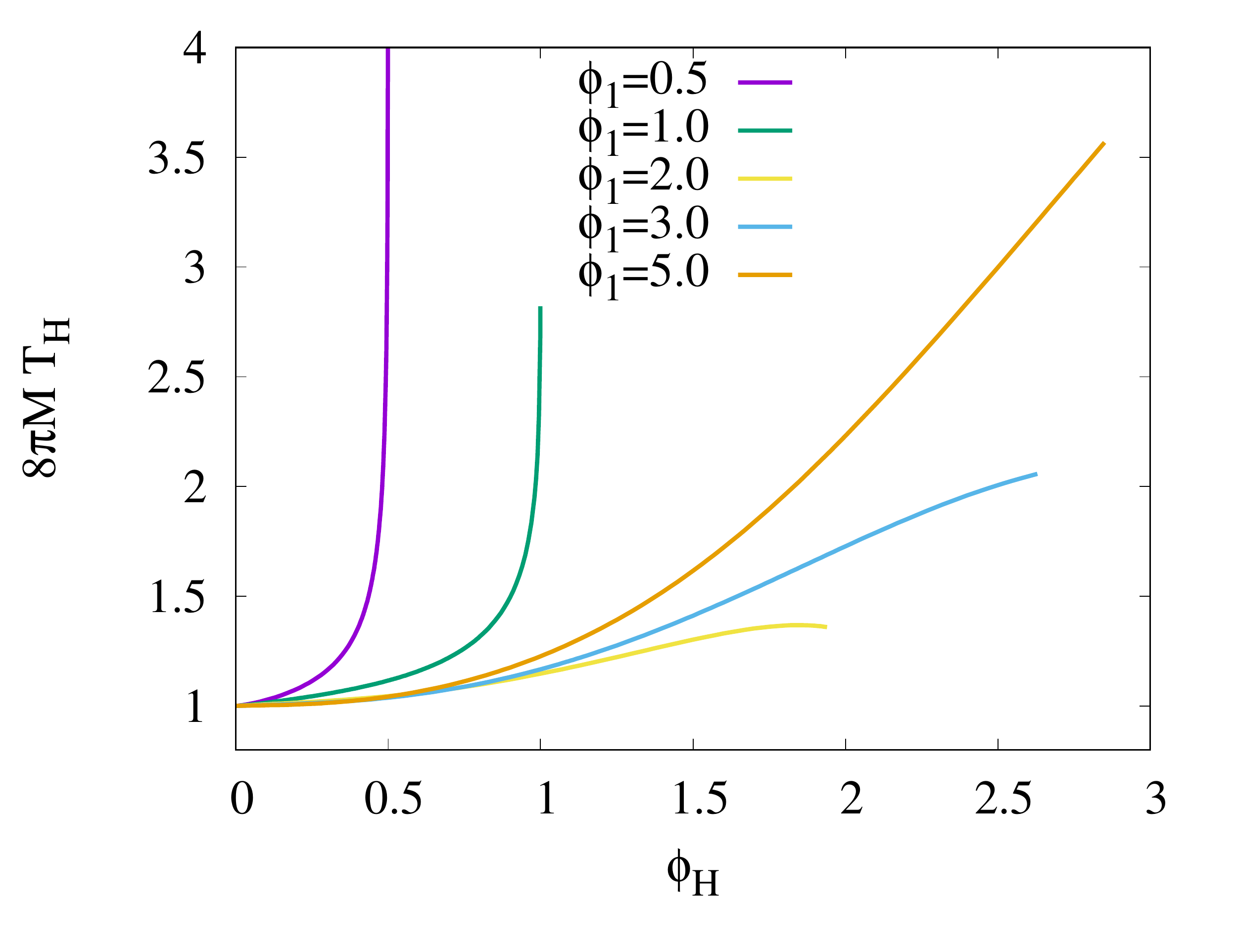}
 }
\caption{The properties of hairy black holes $(r_H=1, V_0=1)$ with several global minimum $\phi_1$ as a function of $\phi_H$: (a) The scaled area of horizon $a_H$. (b) The scaled Hawking temperature $t_H$.}
\label{plot_BH1}
\end{figure}

In summary, this potential $V$ possesses a local maximum and a global minimum which causes $V$ to be negative in some region. The local minimum at zero guarantees that the black hole is asymptotically flat. Moreover, it is possible to repeat the analyisis to show the existence of solitonic solutions, just by changing the lower end of integration (no horizon with a fully regular spacetime).

\subsection{Ordinary Differential Equations (ODEs)}

By substituting Eq. \eqref{line_element} into the Einstein-matter field equation, we obtain a set of second-order and nonlinear ODEs for the metric functions,
\begin{equation}
m' = 2 \pi G r^2 \left( N \phi'^2 + 2 V \right) \,, \quad \sigma' = - 4 \pi G r \phi'^2 \,,    \quad
\left(  e^{- \sigma} r^2 N \phi' \right)' = e^{- \sigma} r^2  \frac{d V}{d \phi} \,,
\end{equation}
where the prime denotes the derivative of the functions with respect to the radial coordinate $r$.

To construct globally regular black hole solutions, we need to know the asymptotic behavour of the functions at the horizon and the infinity. By making the series expansion for the functions at the horizon, the leading terms in the series expansion are given by
\begin{align}
 m(r) &= \frac{r_H}{2}+ m_1 (r-r_H) + O\left( (r-r_H)^2 \right) \,, \\
\sigma(r) &= \sigma_H + \sigma_1   (r-r_H) + O\left( (r-r_H)^2 \right)  \,, \\
 \phi(r) &= \phi_H +  \phi_{H,1}  (r-r_H) + O\left( (r-r_H)^2 \right)  \,,
\end{align} 
where
\begin{equation}
   m_1 = 4 \pi G r^2_H  V(\phi_H)  \,, \quad  \sigma_1 = -  4 \pi G r_H \phi_1 \,, \quad   \phi_{H,1}= \frac{r_H \frac{d V(\phi_H)}{d \phi}}{1-8 \pi G r_H^2 V(\phi_H)}  \,.
\end{equation} 
Here $\sigma_H$ and $\phi_H$ are the values of $\sigma$ and $\phi$ at the horizon. Similarly, the leading terms in the series expansion for scalaron at the origin are given by
\begin{align}
 m(r) &= \frac{4 \pi G V(\phi_c)}{3}  r^3 + O(r^5) \,, \\
\sigma(r) &= \sigma_c  - \frac{\pi G}{9} \left( \frac{d V(\phi_c)}{d \phi} \right)^2 r^4 + O(r^8)  \,, \\
 \phi(r) &= \phi_c +  \frac{1}{6} \frac{d V(\phi_c)}{d \phi}  r^2 + O(r^4)  \,,
\end{align} 
where $\sigma_c$ and $\phi_c$ are the values of $\sigma$ and $\phi$ at the origin, respectively.

\begin{figure}[t!]
\centering
\mbox{
(a)
 \includegraphics[angle =-90,scale=0.33]{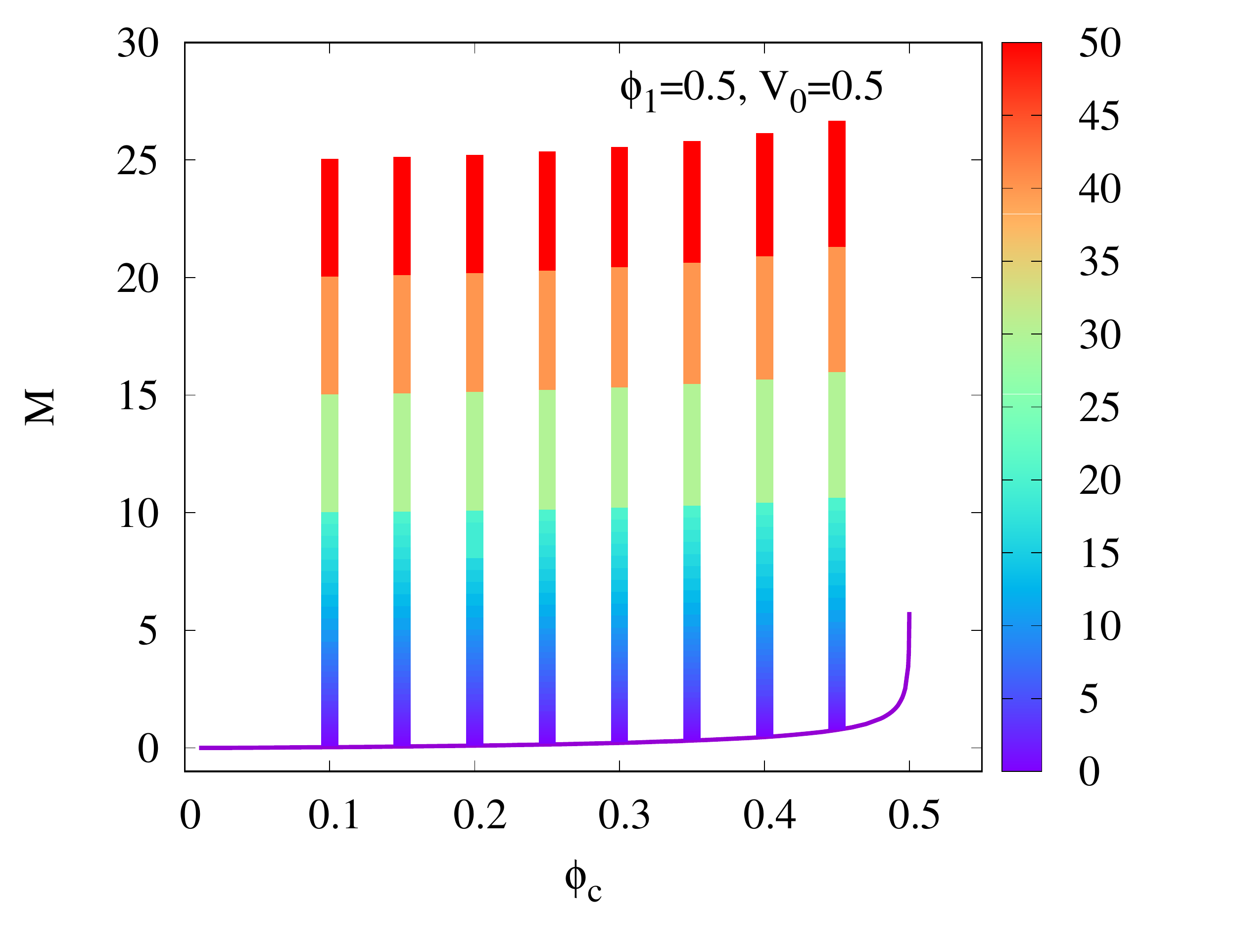}
(b)
 \includegraphics[angle =-90,scale=0.33]{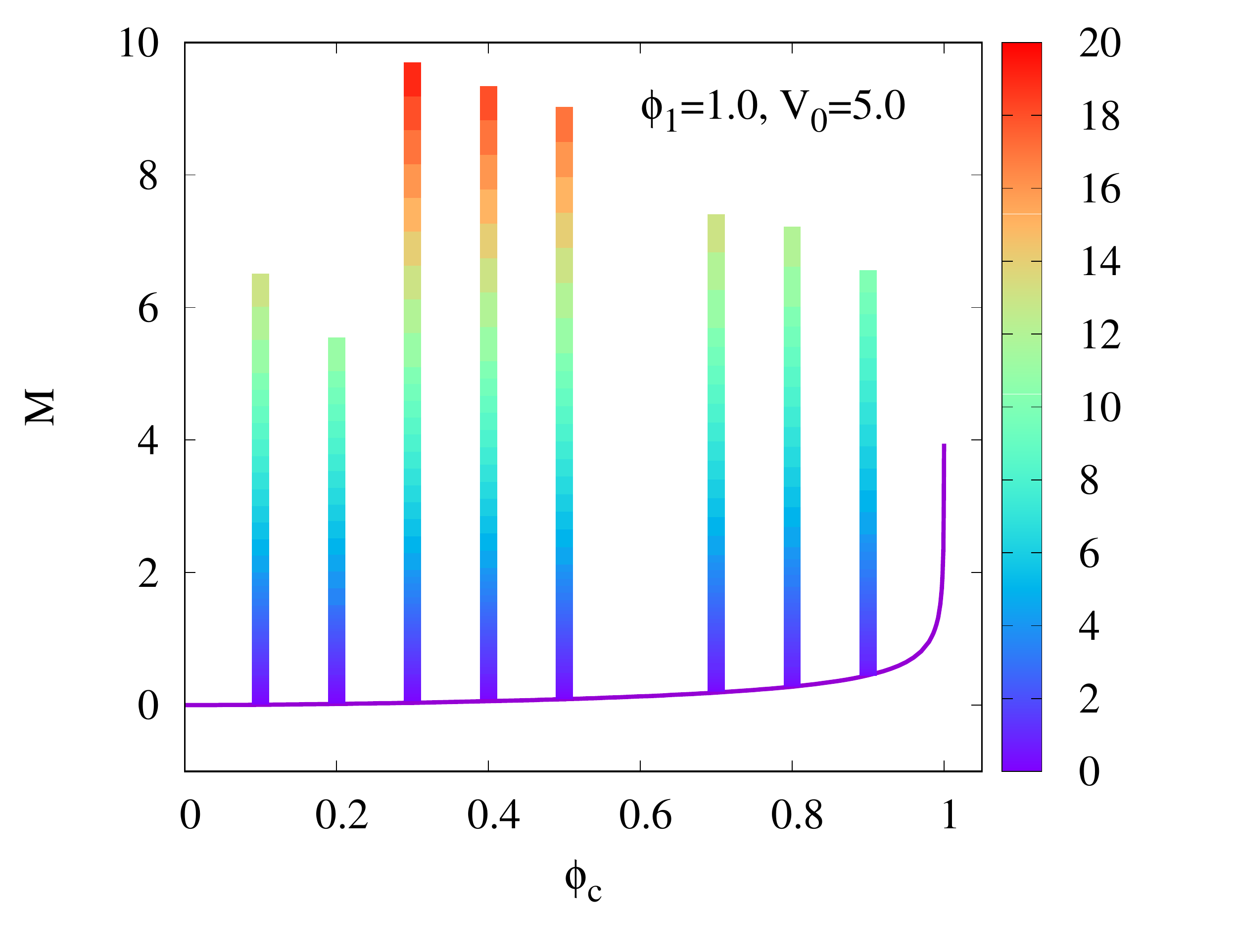}
 }
\mbox{
(c)
 \includegraphics[angle =-90,scale=0.33]{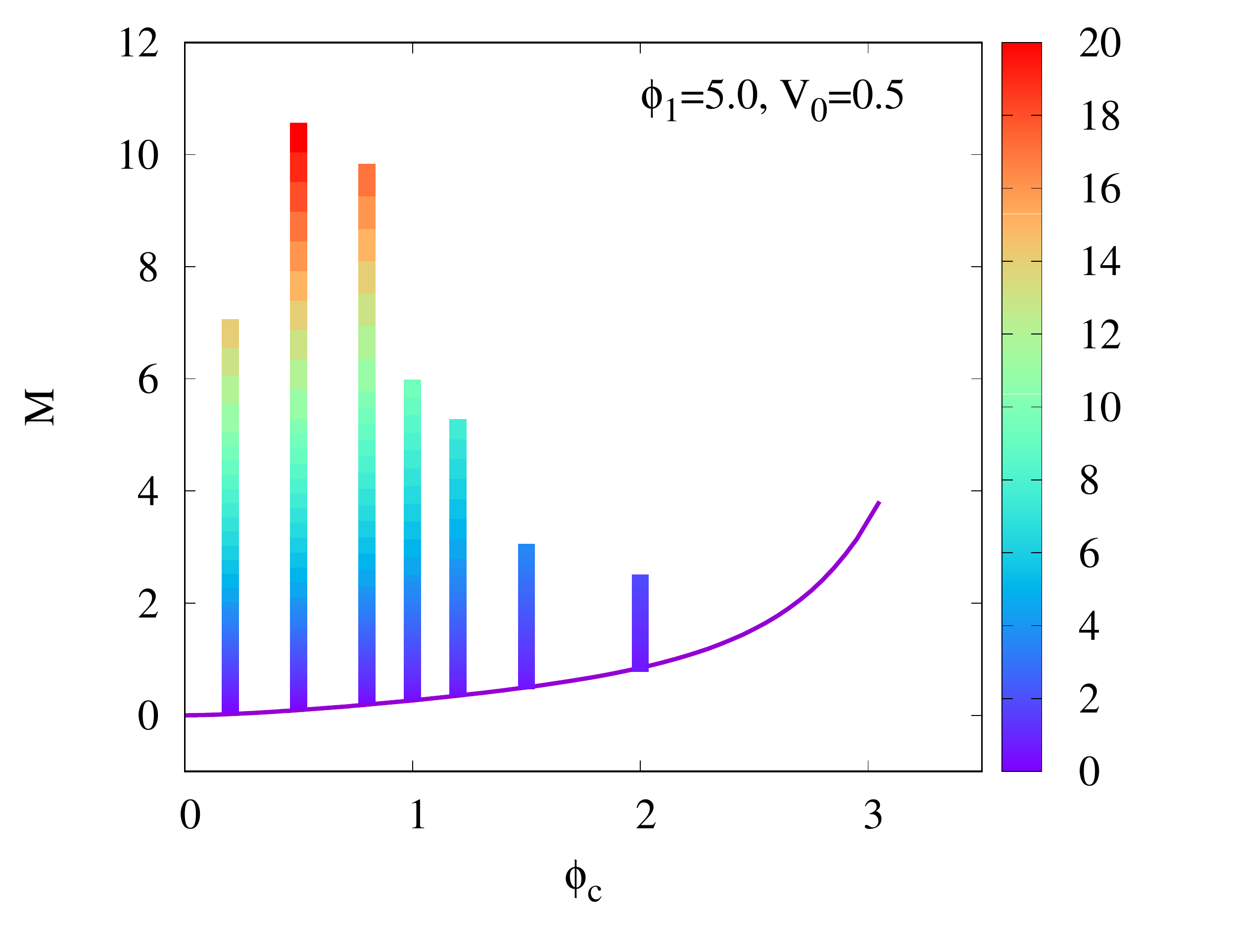}
}
\caption{The mass $M$ as a function of the parameter $\phi_c$. The purple curve corresponds to the scalaron, and the rest of colored lines for sets of black holes with fixed values of $\phi_H$ (in this case $\phi_c=\phi_H$). The colour gradient indicates the size of the horizon radius $r_H$.}
\label{plot_BH2}
\end{figure}

Black holes and scalarons share the same asymptotic expansion of the metric and scalar functions at infinity. As we take $r\to \infty$, if we impose asymptotical flatness and the scalar field to vanish, 
then the leading terms are given by the following expressions
\begin{align}
    m(r) &= M  + \tilde{m}_1 \frac{\exp{(- 2 m_{\text{eff}} r)}}{r} + ...\, , \\
    \sigma(r) &= \tilde{\sigma}_1 \frac{ \exp{(- 2 m_{\text{eff}} r)}}{r}  +   ... \, , \\
    \phi(r) &= \bar{\phi}_{H,1}  \frac{ \exp{(- m_{\text{eff}} r)} }{r} + ... \, ,
\end{align}
where $\tilde{m}_1$, $\tilde{\sigma}_1$ and $\tilde{\phi}_{H,1}$ are constants, $M$ is the total mass of the configuration. 
Note that the effective mass of scalar field is given by $m_{\text{eff}}=\sqrt{V_0 \psi_0 \psi_1}$.

We introduce the following dimensionless parameters, 
\begin{equation}
 r \rightarrow  \frac{r}{\sqrt{4 \pi G}} \,, \quad m \rightarrow  \frac{m}{\sqrt{4 \pi G}} \,, \quad \phi \rightarrow \sqrt{4 \pi G} \phi \,.
\end{equation}
The ODEs are solved by an ODE solver package Colsys which employing the Newton-Raphson method to solve the boundary value problem for a set of nonlinear ODEs by providing the adaptive mesh refinement to generate the solutions to have more than 1000 points with high accuracy and the estimation of errors of solutions \cite{Ascher:1979iha}. We compactify the radial coordinate $r$ by $r=r_H/(1-x)$ for hairy black holes and $r=x/(1-x)$ for scalaron in the numerics. Here we have 5 parameters $(\phi_H,\phi_0,\phi_1, V_0, r_H)$ to describe the hairy black holes and 4 parameters $(\phi_c,\phi_0,\phi_1, V_0)$ to describe the scalaron. To generate the solutions, we fix the global minimum $\phi_1$, the value of $\phi_0$ is determined exactly when the boundary conditions are satisfied with the suitable choices of $\phi_c$ and $\phi_H$ for scalaron and hairy black holes with fixed $r_H$ and $V_0$, respectively. 

\begin{figure}[t!]
\centering
\mbox{
(a)
 \includegraphics[angle =-90,scale=0.33]{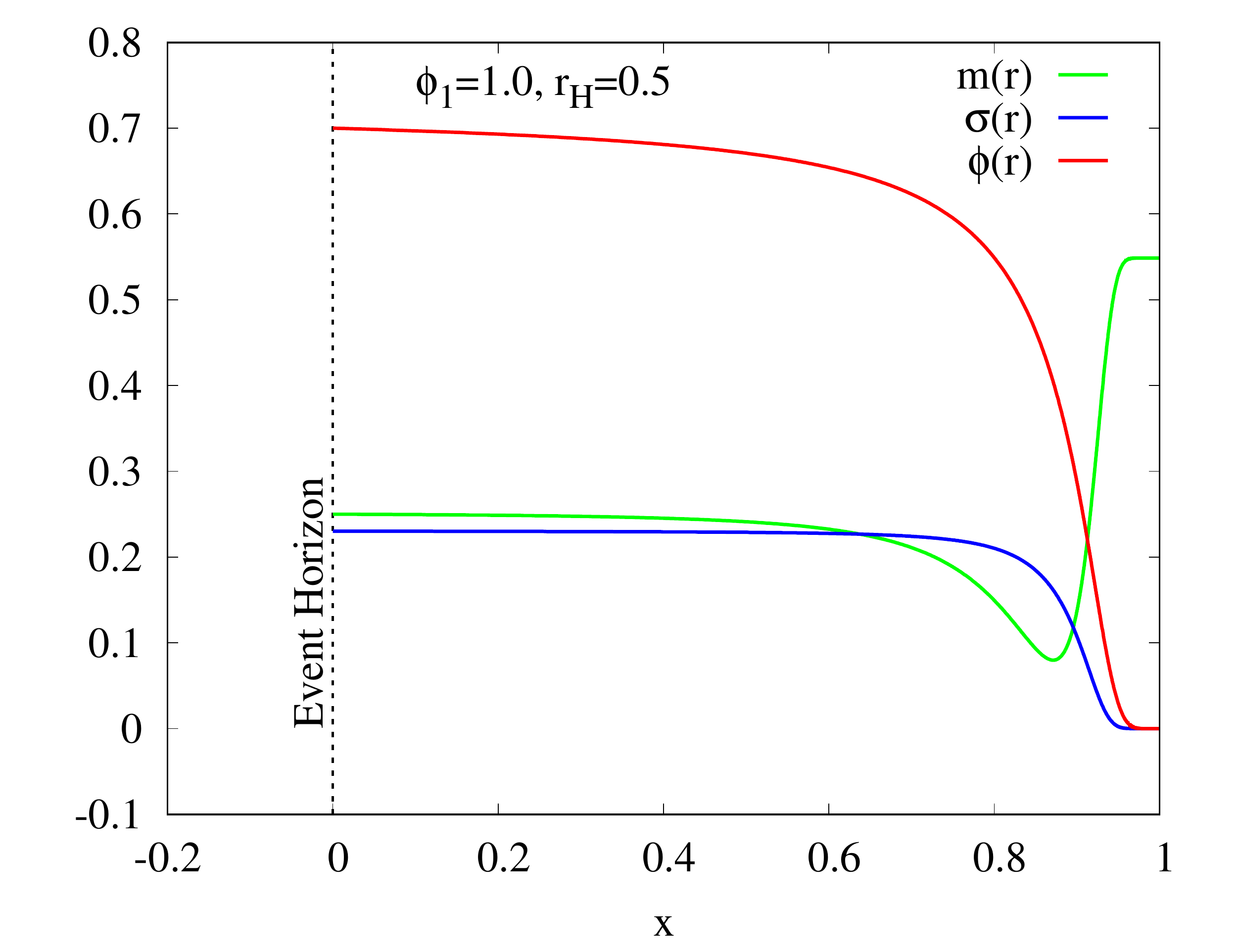}
(b)
 \includegraphics[angle =-90,scale=0.33]{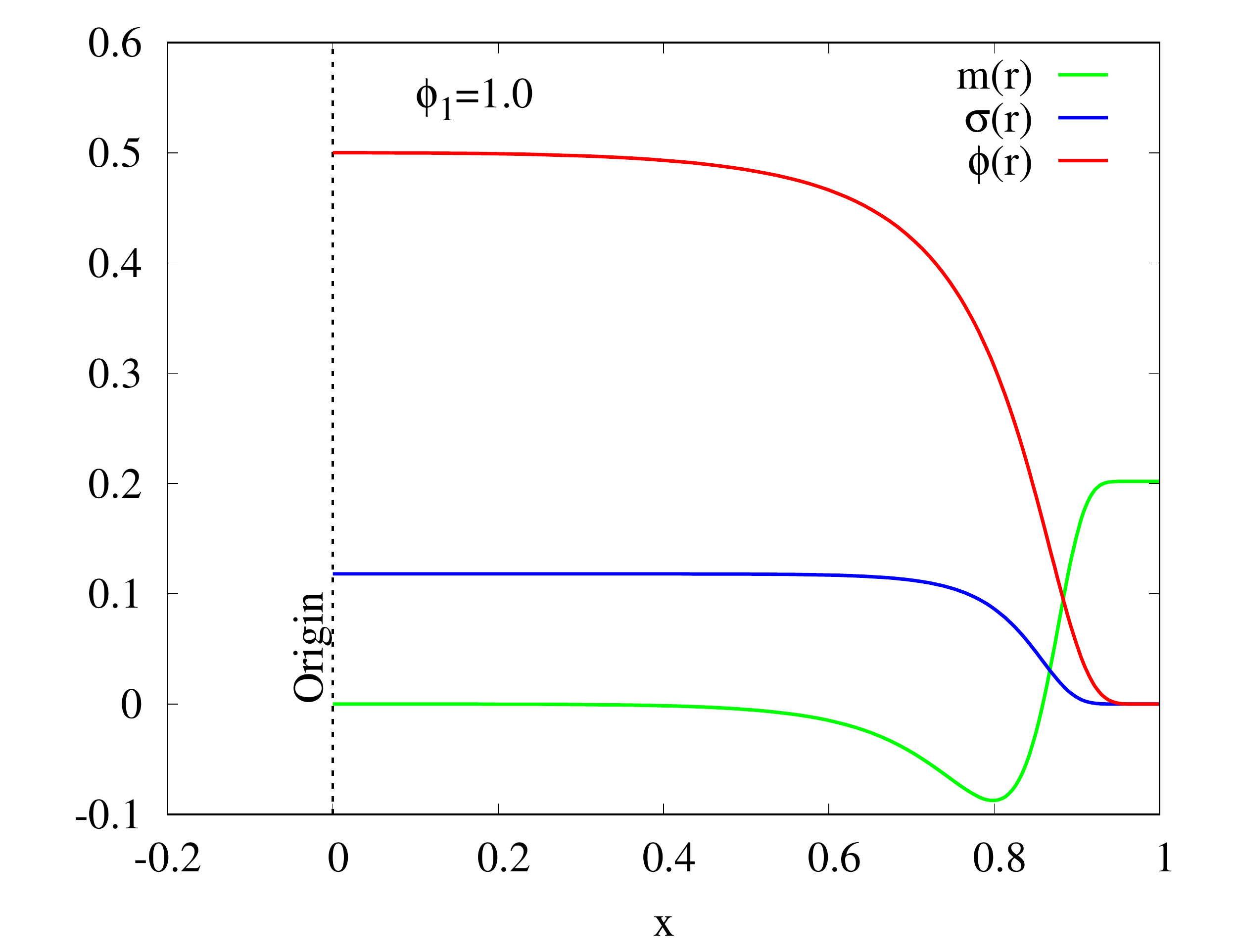}
 }
\caption{Typical profiles of the functions in the compactified coordinate $x$ for (a) Black holes. (b) Scalaron. }
\label{plot_profile}
\end{figure}

If we consider a hairy black hole, our solutions require a non-vanishing gradient of the scalar field at the event horizon. Therefore, if we maximally extend the causal structure over the Einstein-Rosen bridge (left of Fig. \ref{plot_causal}), one necessarily sees a cusp of the scalar field (black dashed curve) inside the horizon. However, it is fair to say that our solutions satisfy the regularity condition at the event horizon, and hence, there is a natural extension of the solution beyond the event horizon up to the singularity. Thus, if we restrict the situation that our solution is formed from astrophysical processes, e.g., gravitational collapses, we can restrict our solutions above a timelike hypersurface (blue dashed curve), e.g., that denotes a star surface. So, the final physically sensible causal structure becomes right of Fig. \ref{plot_causal}.

\section{Properties of the Scalar Hairy Black Holes} \label{sec:prop}

We are interested with the Hawking temperature $T_H$ and area of horizon $A_H$ of these black holes,
\begin{equation}
T_H = \frac{1}{4 \pi} N'(r_H) e^{-\sigma_H} \,, \quad A_H = 4 \pi r^2_H \,,
\end{equation}
where $\sigma_H$ is the value of function $\sigma$ at the horizon. For the convenience of comparing our black hole solution with other known solutions, we introduce the following reduced quantities at the horizon of the black holes,
\begin{equation}
 a_H = \frac{A_H}{16 \pi M^2} \,, \quad t_H = 8 \pi T_H M \,.
\end{equation}

The Ricci scalar $R$ and Kretschmann scalar $K$ for the black hole spacetime are given by
\begin{align}
 R &= -N'' + \frac{3 r \sigma'-4}{r} N' + \frac{2 \left( 2 r N \sigma' - N +1 + r^2 N\sigma'' - r^2 N \sigma'^2  \right)}{r^2} \,, \\
 K &= \left(  3 \sigma' N' + 2 N \sigma'' - N'' -2 N \sigma'^2  \right)^2    +  \frac{2}{r^2} \left( N'-2 N \sigma'  \right)^2 + \frac{2 N'^2}{r^2} + \frac{4 (N-1)^2}{r^4}  \,.
\end{align}
With the series expansion of functions at the horizon, the $R$ and $K$ are finite with the leading order,
\begin{align}
  R &=  -\frac{2 m_1 \left( 3 r_H \sigma_1 -2 \right)}{r^2_H} + \frac{3 \sigma_1 + 4 m_2}{r_H} +  O\left( (r-r_H) \right)  \,, \\
  K &=  \frac{16 m^2_2}{r^2_H} - \frac{8 \left(  -2 + 6 m_1 \sigma_1 r_H + 4 m_1 - 3 \sigma_1 r_H  \right)}{r^3_H}   \nonumber \\
& \quad + \frac{12-32 m_1 + 48 m^2_1 \sigma_1 r_H + 36 m^2_1 \sigma^2_1 r^2_H + 32 m^2_1 + 9 \sigma^2_1 r^2_H + 12 \sigma_1 r_H - 36 \sigma^2_1 r^2_H m_1 - 48 m_1 \sigma_1 r_H}{r^4_H}   +  O\left( (r-r_H) \right)  \,.
\end{align}

\begin{figure}
\centering
 \mbox{
 (a)
 \includegraphics[angle =-90,scale=0.33]{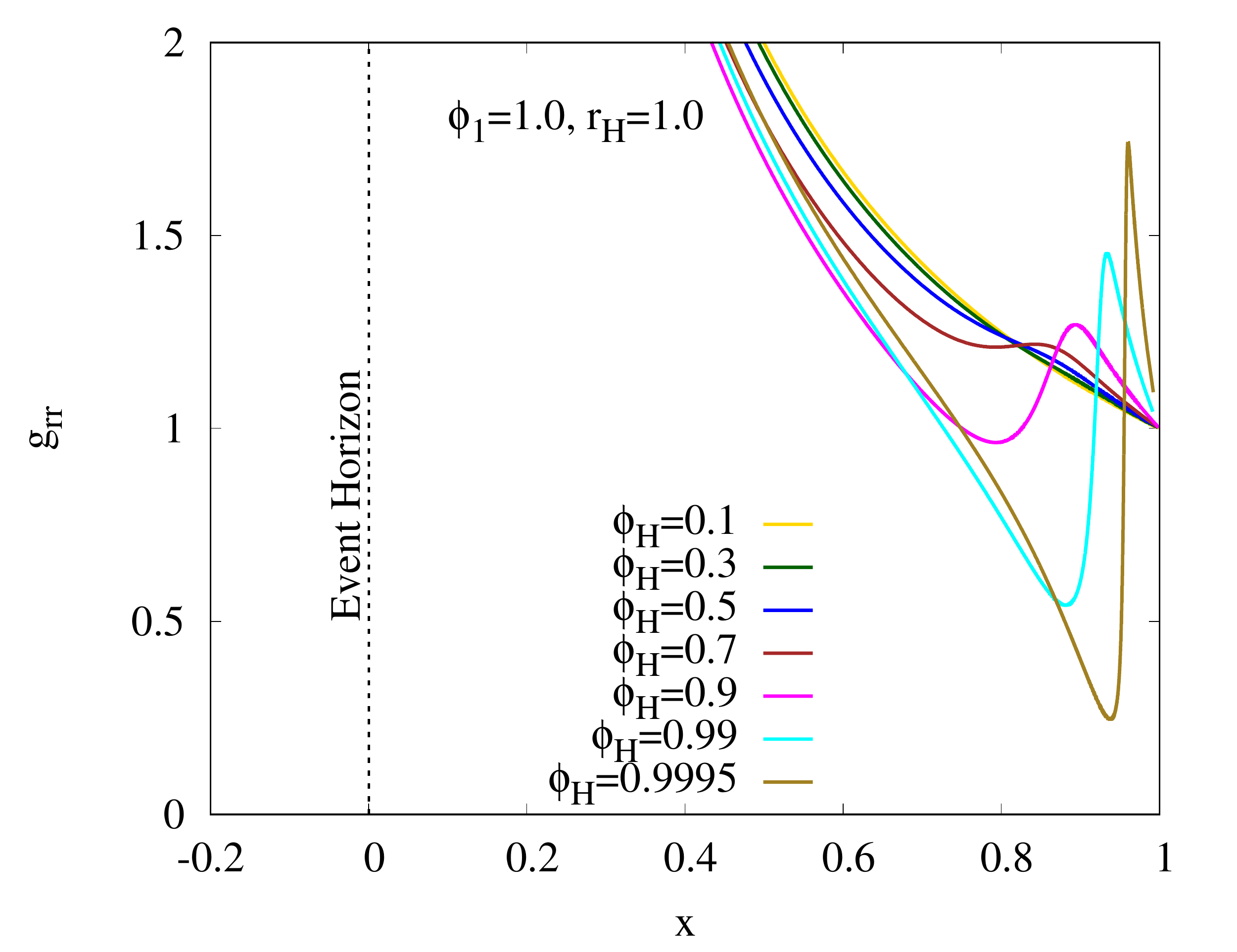}
(b)
 \includegraphics[angle =-90,scale=0.33]{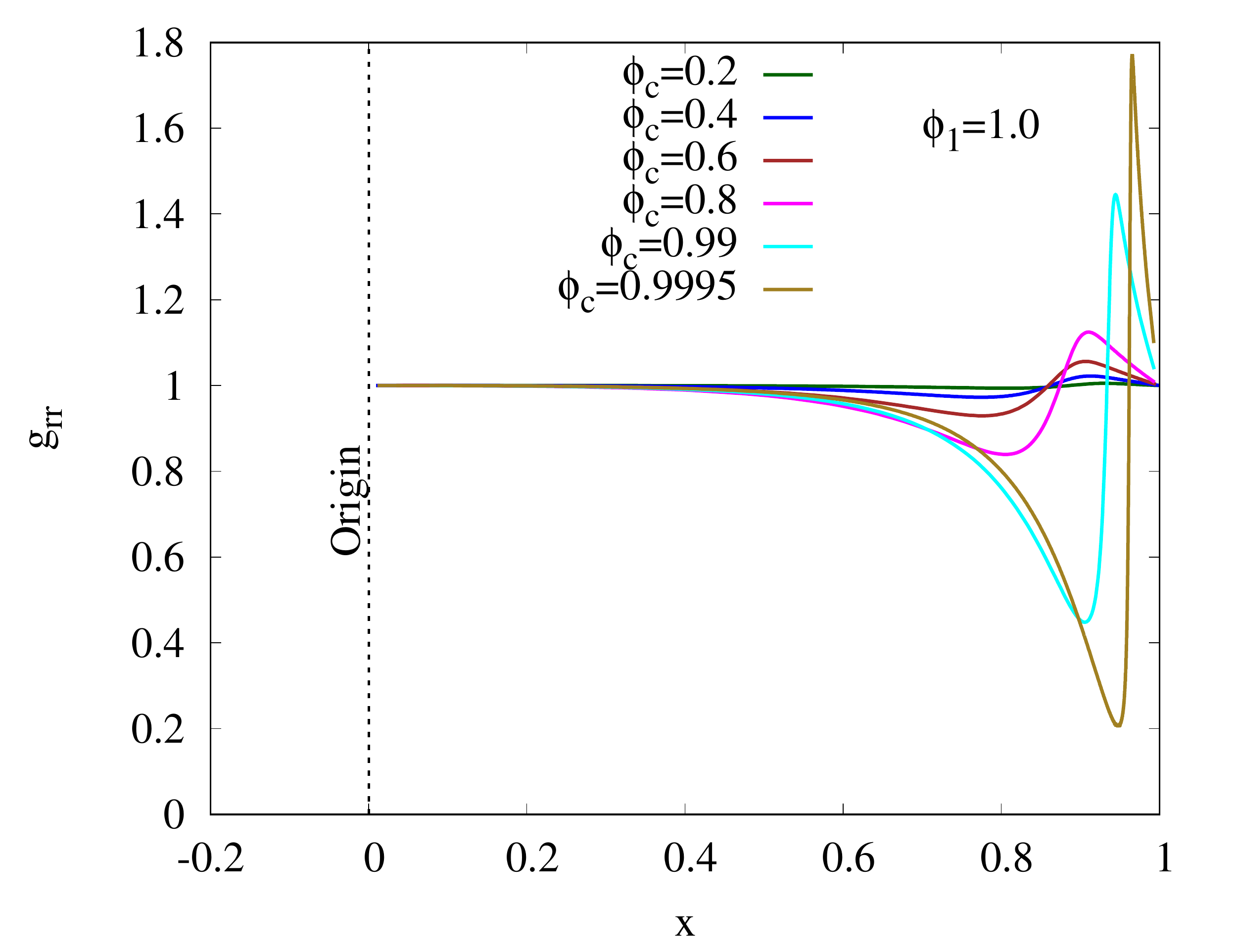}
 }
\caption{The metric component $g_{rr}$ for a) hairy black hole b) scalaron in the compactified coordinate $x$.}
\label{plot_grr}
\end{figure}

\section{The Radial Perturbation}\label{sec:lin}

In the case of radial perturbation, we perturb the background metric and scalar field, respectively as \cite{Blazquez-Salcedo:2020nhs}
\begin{eqnarray}
 ds^2 &=& - N(r) e^{-2 \sigma(r)} \left[  1 + \epsilon e^{-i \omega t} F_t(r)  \right] dt^2 + \frac{1}{N(r)} \left[ 1+ \epsilon e^{-i \omega t} F_r(r)   \right] dr^2 + r^2 \left( d \theta^2+\sin^2 \theta d\varphi^2 \right) \,, \\
 \Phi &=& \phi(r) + \epsilon  \Phi_1 (r) e^{-i \omega t} \,,
\end{eqnarray}
where $ F_t(r)$,  $F_r(r)$ and $\Phi_1 (r)$ are the small perturbations to the non-perturbed solutions. 

By substituting the above Ansatz to the Einstein equation and KG equation, we obtain a set of ODEs for the perturbation functions, 
\begin{eqnarray}
 F_r &=& 8 \pi G r \Phi_1 \phi'   \,, \label{ODEper1} \\
 F'_t &=& - F'_r + 16 \pi G r \Phi'_1 \phi'   \,, \label{ODEper2} \\
\Phi''_1 &=& \left(  \sigma' - \frac{N'}{N} -\frac{2}{r} \right)  \Phi'_1 + \left(  \frac{1}{N}  \frac{d^2 V}{d \phi^2}  - \omega^2 \frac{e^{2 \sigma}}{N^2}  \right)  \Phi_1 + \frac{F_r}{N}\frac{d V}{d \phi} + \frac{ F'_r - F'_t }{2} \phi'  \,. \label{ODEper3}
\end{eqnarray}
We observe that only Eq. \eqref{ODEper3} is independent because other ODEs (Eqs. \eqref{ODEper1} and \eqref{ODEper2}) are dependent. Hence, we transform $\Phi''_1 $ to a Schr\"odinger-like master equation by defining $Z(r) = r \Phi_1(r)$,
\begin{equation}
 \frac{d^2 Z}{d r_{*}^2} + \left( \omega^2 - V_R(r) \right) Z =0 \,,
 \label{Z_radial}
\end{equation}
with the effective potential $V_R(r)$,
\begin{equation}
    V_R(r) = N e^{-2 \sigma} \left[ \frac{N}{r} \left( \frac{N'}{N} - \sigma'  \right)  - 8 \pi G r N \phi'^2 \left(  \frac{N'}{N} + \frac{1}{r} - \sigma' \right) + 16 \pi G r \phi' \frac{d V}{d \phi}  + \frac{d^2 V}{d \phi^2} \right]   \,.
 \label{V_R}
\end{equation}
The tortoise coordinate $r_*$ is 
\begin{equation}
 \frac{d r_*}{d r} = \frac{e^{\sigma}}{N}  \,.
\end{equation}
The perturbation $Z$ is unstable when $\omega^2<0$ where the perturbation grows exponentially with time. In the compactified coordinate $x$, the effective potential $V_R$ has the following expansions at the origin and horizon, respectively,
\begin{align}
 \text{Scalaron}: V_{R}(0) &= \tilde{V}_0 + \tilde{V}_1 x + \tilde{V}_2 x^2 + O(x^3) \,, \\
 \text{Black Hole}: V_{R}(0) &=  \hat{V}_1 x + \hat{V}_2 x^2 + O(x^3) \,,
\end{align} 
where $\tilde{V}_i$ and $\hat{V}_i$ are constants that depend on the parameters of the background solution.

Note that Eq.~\eqref{Z_radial} is an eigenvalue problem, hence we compute the radial mode numerically by using COLSYS to solve it  with $\omega^2$ as the eigenvalue. For black holes, we impose that the first-order derivative of the perturbation function vanishes at the boundaries, $\partial_{r}Z(r_H)=\partial_{r}Z(\infty)=0$. In the case of the scalaron, we impose that the perturbation function vanishes at the boundaries. In the numerics, we introduce an auxiliary equation $\frac{d}{d r} \omega^2=0$, that allows us to impose an additional condition $Z(r_p)=1$ at some point $r_p$, which typically lies in the middle of the horizon/origin and infinity. This allows us to obtain a nontrivial and normalizable solution for $Z$, since Eq.~\eqref{Z_radial} is homogeneous. The eigenvalue $\omega^2$ is found automatically when $Z$ satisfies all the asymptotic boundary conditions.

\begin{figure}
\centering
\mbox{
(a)
 \includegraphics[angle =-90,scale=0.33]{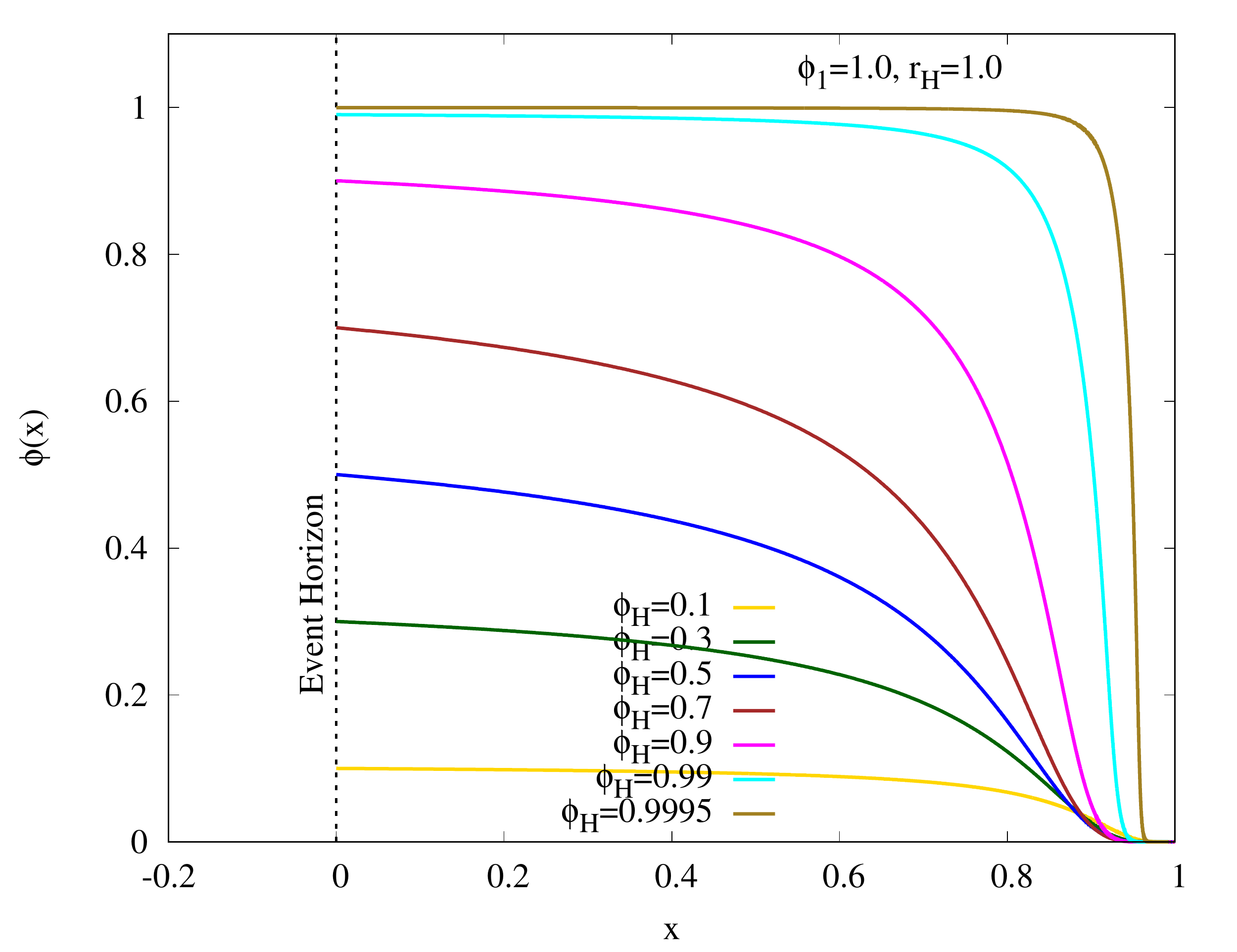}
(b)
 \includegraphics[angle =-90,scale=0.33]{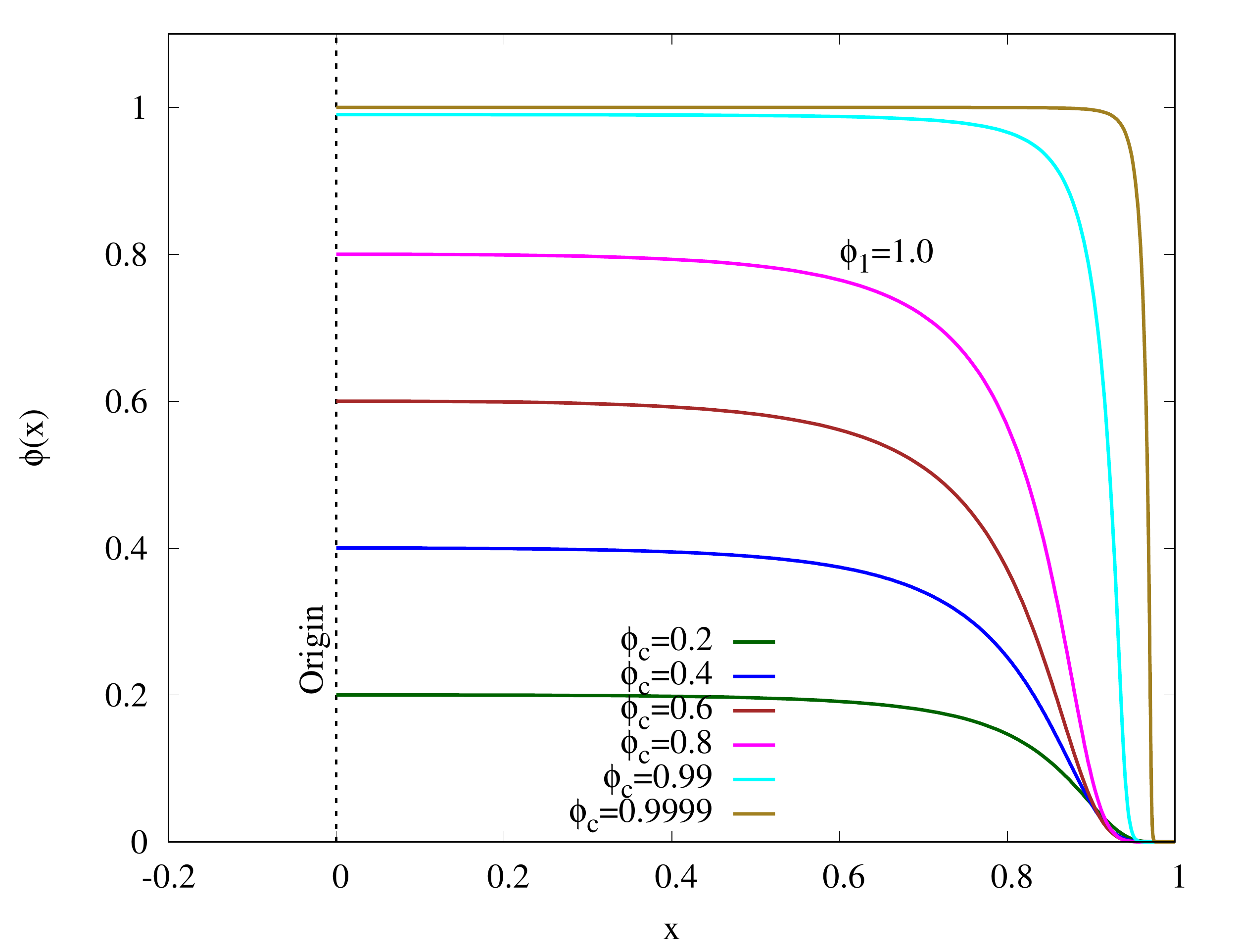}
 }
\caption{Profiles of scalar field $\phi$ for a) hairy black hole b) scalaron in the compactified coordinate $x$.}
\label{plot_dil}
\end{figure}

\section{Results and Discussions}\label{sec:res}

\subsection{General properties and domain of existence.}

By fixing several values of global minimum $\phi_1$, we exhibit the properties of hairy black holes which are the reduced area of horizon $a_H$ and reduced Hawking temperature $t_H$ in Fig. \ref{plot_BH1}. The purpose to introduce the reduced quantities is to compare our hairy black holes with the known solution which is the Schwarzschild black hole in this case. Recall that both $a_H$ and $t_H$ are unity for Schwarzschild black hole. By increasing the value of scalar field at the horizon $\phi_H$ from zero, a branch of hairy black hole solutions emerges from the Schwarzschild black hole. As $\phi_H$ increases, $a_H$ decreases from unity and $t_H$ increases from unity. When $\phi_H \rightarrow \phi_1$, $a_H$ decreases to zero and $t_H$ increases very sharply for $\phi_1=0.5, 1.0$. 
We couldn't generate configurations for $\phi_H = \phi_1$, the solutions becoming singular as we reach this value of the parameter. Note that when $\phi_H = \phi_1$, the scalar field sits exactly at the true vacuum $\phi_1$ of the potential and the tunneling effect does not occur. 

For higher values of $\phi_1$ (i.e. $\phi_1=2.0, 3.0, 5.0$), the reduced area $a_H$ also decreases almost to zero but the reduced temperature $t_H$ remains finite. Again, the solutions should become sick in the limit $\phi_H = \phi_1$, but numerically reaching this limit is more complicated, since the code stopping to work for values of $\phi_H$ slighlty below $\phi_1$ (for $\phi_1=2.0, 3.0$) and $\phi_H$ far below $\phi_1$ (for $\phi_1=5.0$). 

\begin{figure}
\centering
\mbox{
(a)
 \includegraphics[angle =-90,scale=0.33]{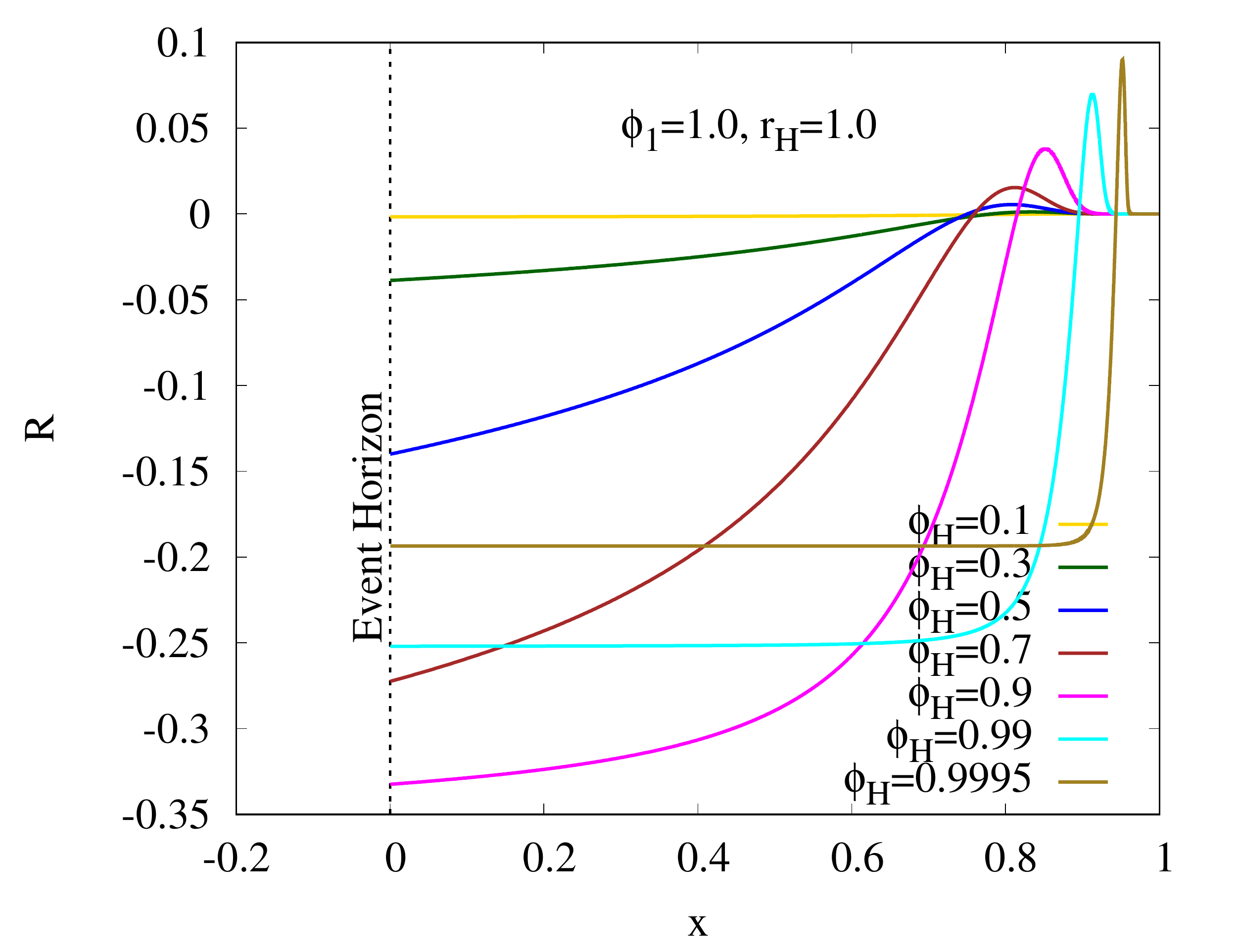}
(b)
 \includegraphics[angle =-90,scale=0.33]{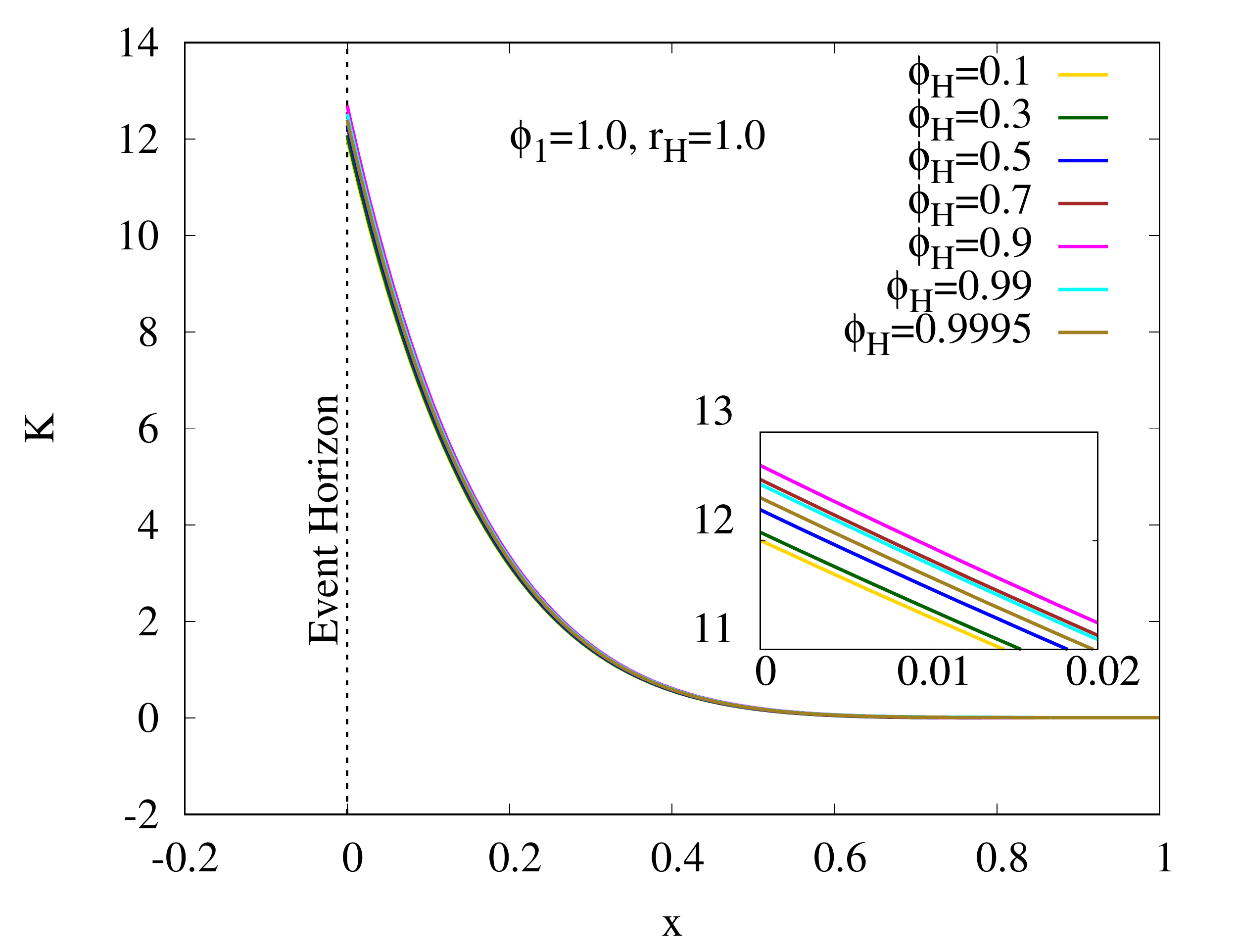}
 }
\caption{(a) The Ricci scalar and (b) The Kretschmann scalar for a hairy black hole in the compactified coordinate $x$.} %
\label{plot_Ricci_Kretschmann}
\end{figure}

In Fig. \ref{plot_BH2} we show the relation between the mass of hairy black holes and scalarons with the value of $\phi_c$. The scalaron solutions (purple curve) emerge from the Minkowski spacetime, where the mass is zero and the scalar field vanishes. As the value of scalar field at the origin $\phi_c$ increases from zero, the mass of scalaron also increases. When $\phi_c \rightarrow \phi_1$, the mass of scalaron increases very sharply. 

The colour bars in the Fig. \ref{plot_BH2} represent the size of the black hole horizon, which indicate that the mass of scalaron are connected with hairy black holes in the small horizon limit $(r_H \rightarrow 0)$. Thus, the hairy black holes possess the scalaron as the limiting configuration in the small horizon limit. Analogous to the hairy black holes, for $\phi_1=0.5, 1.0$, the scalarons do not exist when $\phi_c = \phi_1$ since the scalar field $\phi_c$ sits exactly at the true vacuum $\phi_1$. For large $\phi_1$, e.g., $\phi_1=5.0$, the scalaron solutions are more difficult to generate numerically, and we cannot reach values of $\phi_c$ too close to $\phi_1$.  

We exhibit the typical profiles of solution in the compactified coordinate $x$ for the hairy black holes and scalaron in Fig \ref{plot_profile}. Both compact objects show the similar pattern for the functions. As the value of the scalar field at the horizon or the origin becomes closer to $\phi_1$, these solutions become closer to bubbles of true vacuum surrounded by the false vacuum. 
We observe the solutions have almost constant functions in the bulk, corresponding to (almost) the global minimum of the potential, and thus the true vacuum $\phi_1$. Moving away from the horizon, the solutions develop a sharp boundary at some intermediate region of the spacetime, where the functions rapidly change to another set of almost constant functions. These region corresponds to the imposed false vacuum $(a=0)$ at infinity, where the scalar sits in the local minimum. 

Moreover, the mass function $m(x)$ possesses a global minimum and can be strictly non-positive (see Fig \ref{plot_profile}(b)), which indicates the violation of the energy conditions. We observe that the global minimum of $m(x)$ decreases very sharply as $\phi_H$ increases to the limit value.
This sharp behaviour of the $m(r)$ function propagates into the metric component $g_{rr}$, that we display in Fig. \ref{plot_grr}.
The global minimum of $m(x)$ gives rise to the global maximum of $N(x)$, since $N(x)=1-2m(x)(1-x)/r_H$. Thus, the global maximum of $N(x)$ also gives rise to the global minimum of $g_{rr}$ since $g_{rr}=1/N(x)$. 

To further illustrate how in the limit the spacetime is divided in two different regions dominated by the two possible vacua, in Fig. \ref{plot_dil} we show the profiles of scalar field for hairy black holes and scalarons. As $\phi_H$ and $\phi_c$ increase to $\phi_1$, the scalar field profile becomes closer and closer to a step function: the bulk region possess the value of the true vacuum $\phi_1$, while the exterior region possess the value of the false vacuum $(a=0)$.

The Ricci scalar $R$ and Kretschmann scalar $K$ for a hairy black hole with $\phi_1=1.0$ are shown in Fig. \ref{plot_Ricci_Kretschmann}(a) and (b), respectively. The Ricci scalar at the horizon decreases from zero to some negative values when $\phi_H$ increases. This indicates that the bulk of hairy black hole possesses negative curvature. From the horizon to a point where the functions have a sharp boundary, the profile of $R$ increases from negative value to zero, then to positive value and then fall to zero again at the infinity. However, the Krestschmann scalar is positive at the horizon, 
and its profile decreases monotonically to zero at the infinity.

\subsection{Spherical stability.}

\begin{figure}[t!]
\centering
\mbox{
(a)
 \includegraphics[angle =-90,scale=0.33]{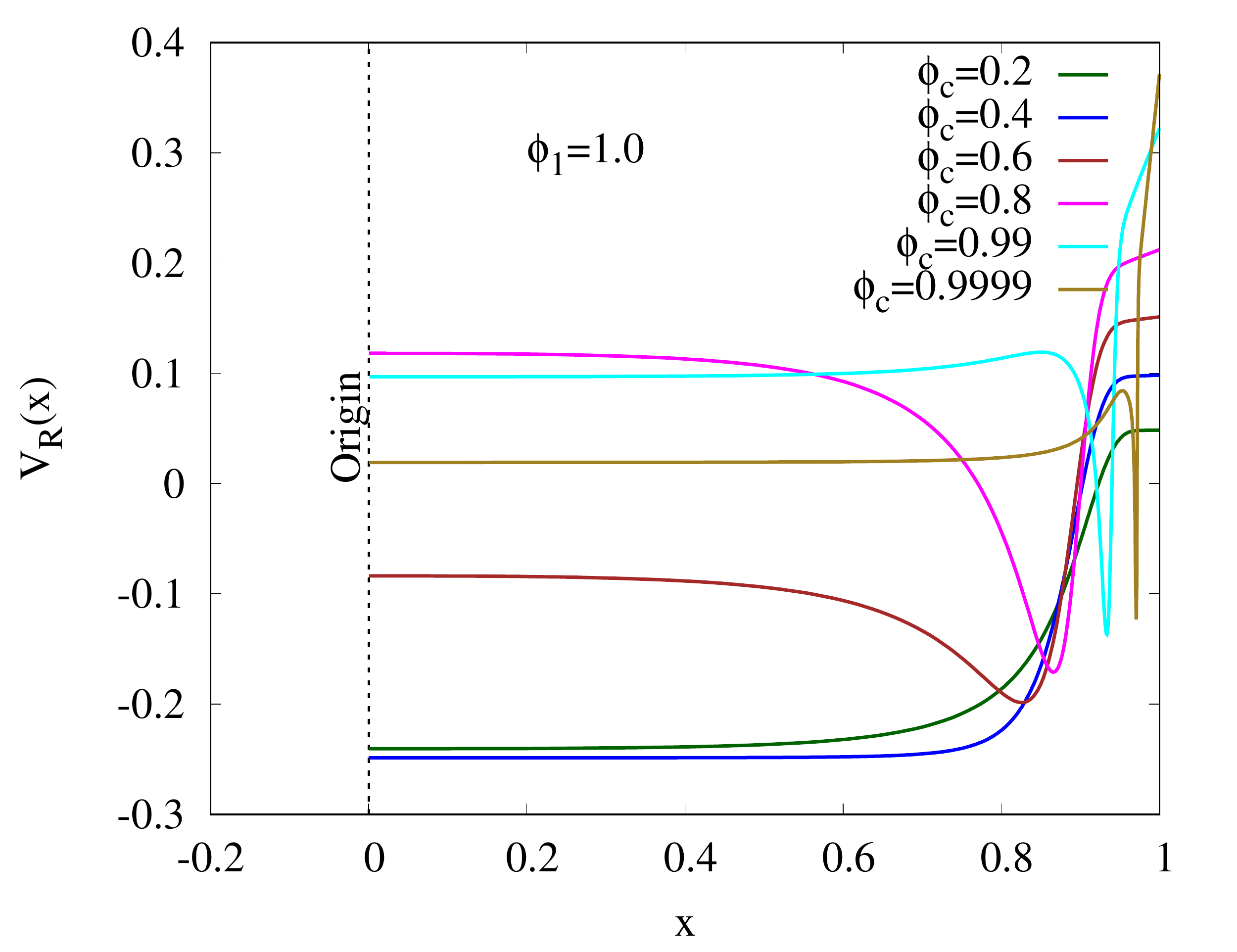}
(b)
 \includegraphics[angle =-90,scale=0.33]{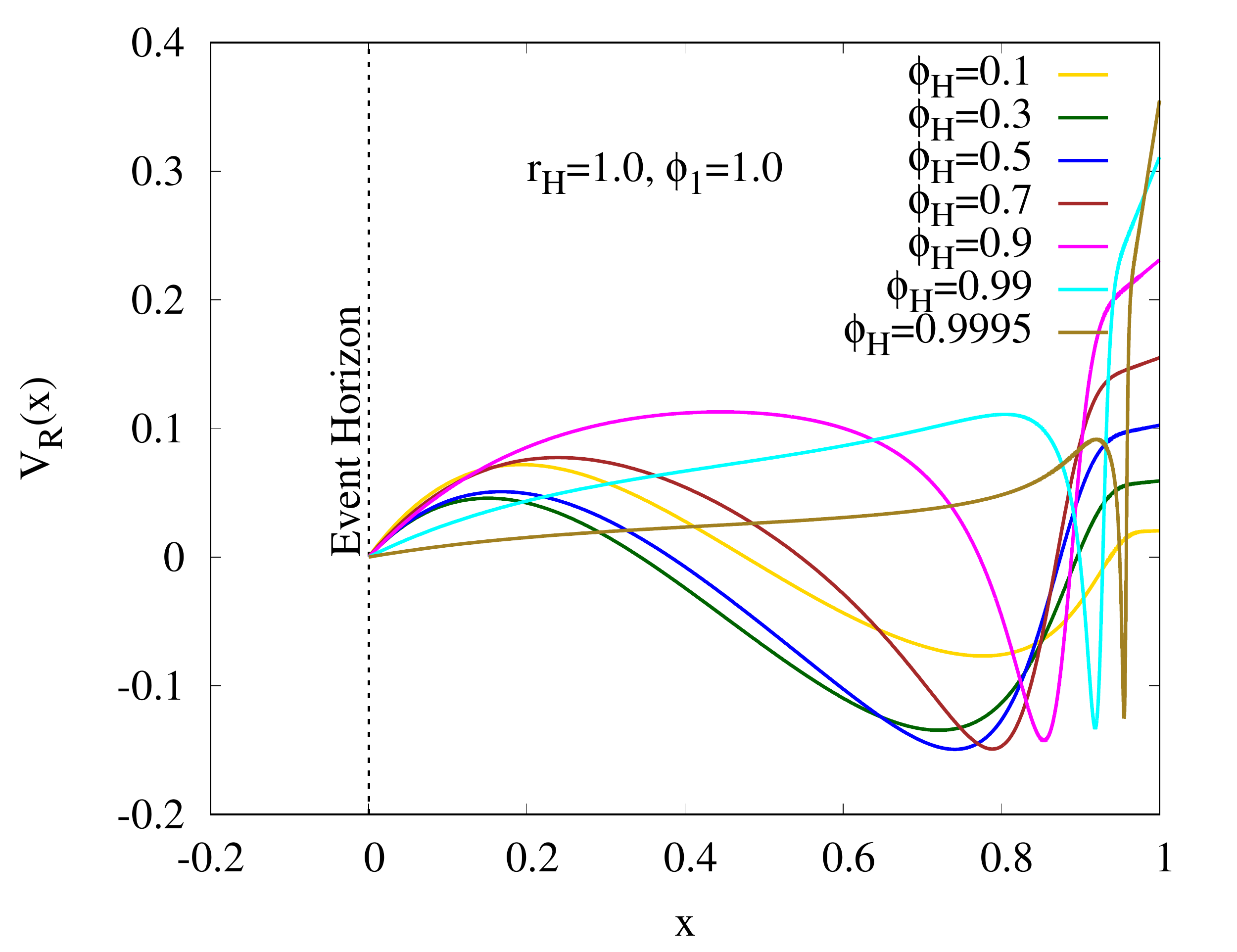}
 }
\caption{Typical profile of the effective potential $V_R(x)$ in the compactified coordinate $x$ for (a) Scalaron (b) Black holes. In colour we plot the potential for different values of the $\phi_c$ parameter.
}
\label{plot_rad_pot}
\end{figure}

In Fig. \ref{plot_rad_pot}, we show the effective potential $V_R$ in the compactified coordinate for the scalaron and hairy black holes. $V_R$ possesses finite value at the origin for scalaron but $V_R$ is zero for the hairy black holes. In this figure it can be seen that the potential possesses always a negative region for both types of objects, indicating the existence of a instability. 

\begin{figure}[t!]
\centering
\mbox{
(a)
 \includegraphics[angle =-90,scale=0.33]{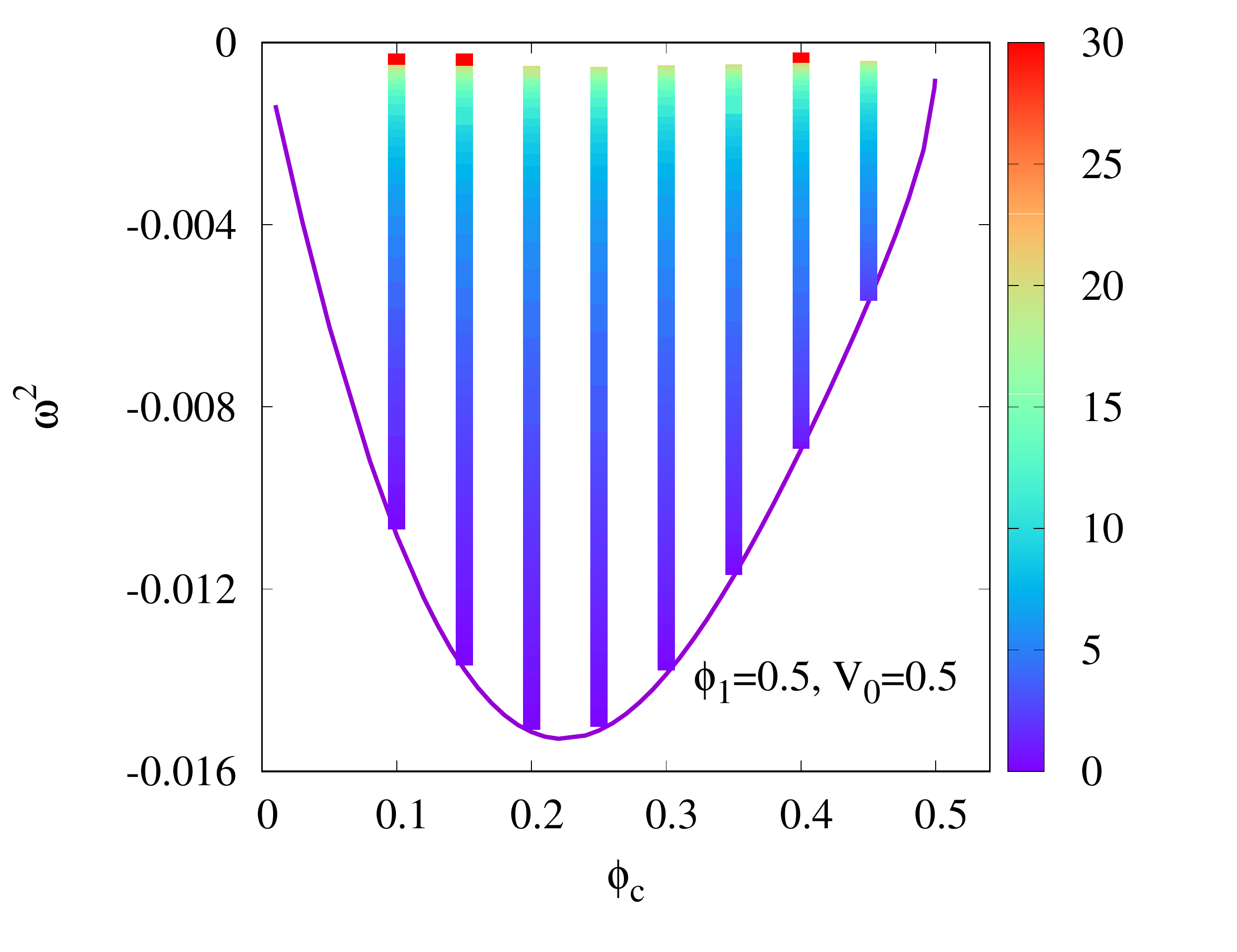}
(b)
 \includegraphics[angle =-90,scale=0.33]{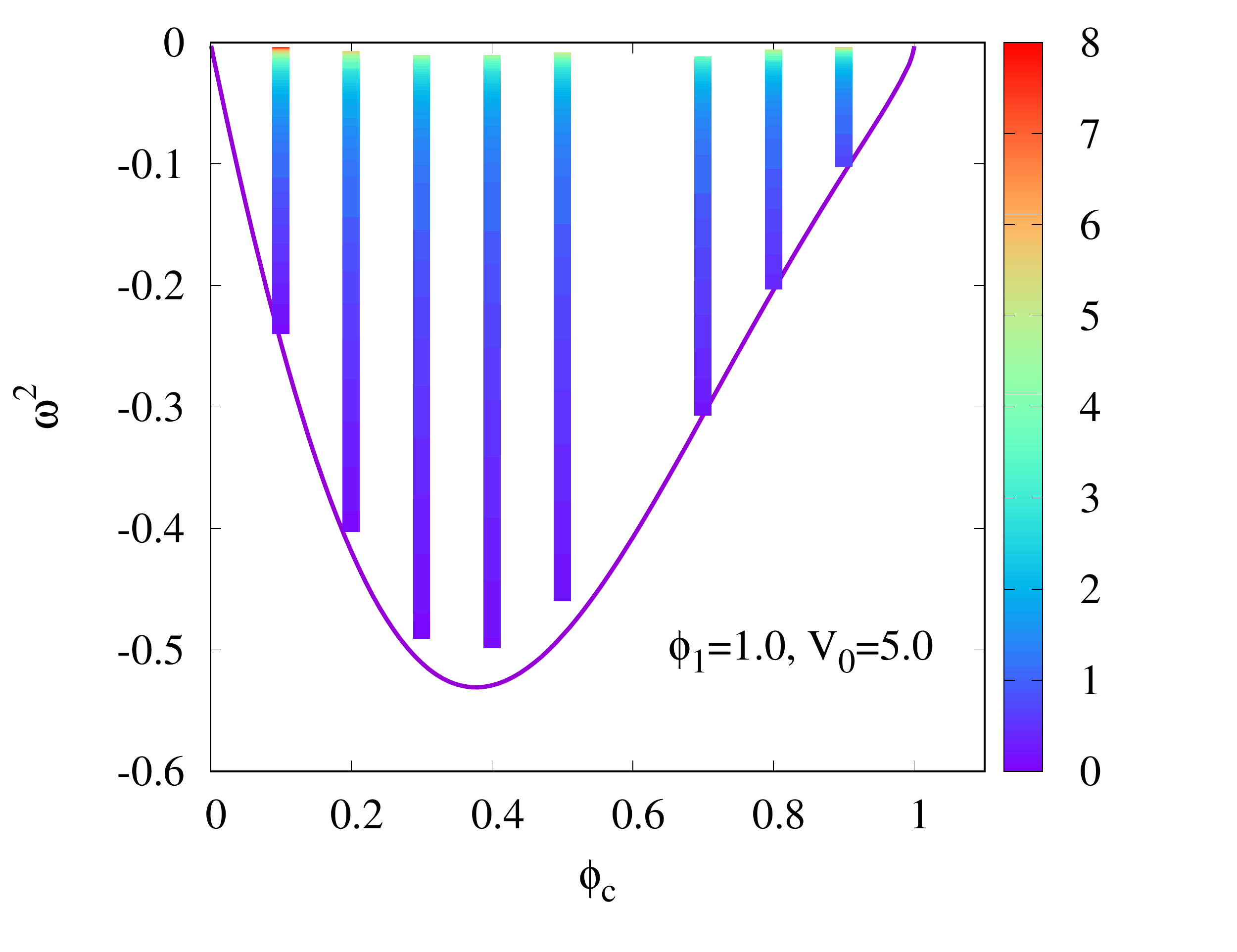}
 }
\mbox{
(c)
 \includegraphics[angle =-90,scale=0.33]{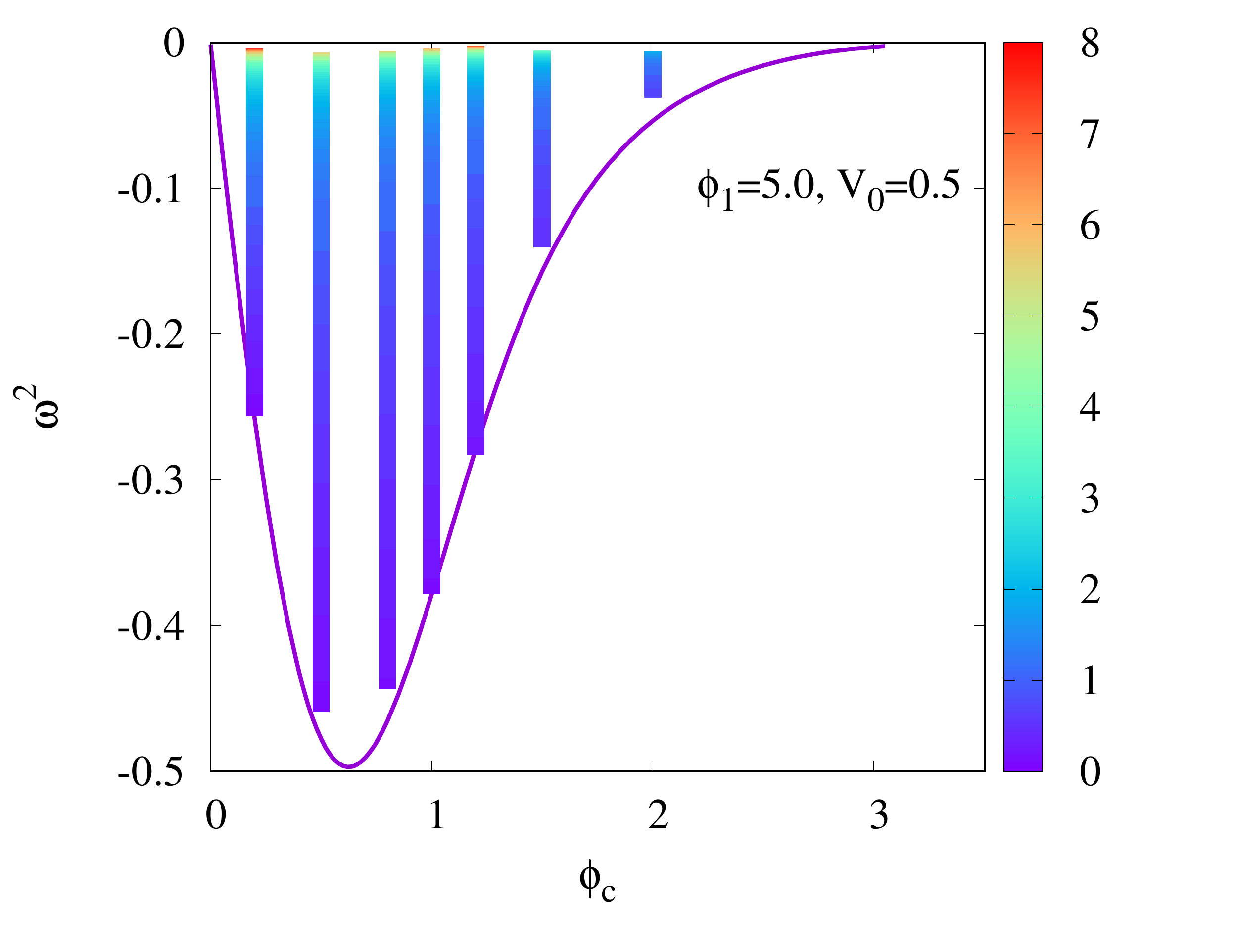}
}
\caption{
The eigenvalue $\omega^2$ as a function of the parameter $\phi_c$. The purple curve corresponds to the scalaron, and the rest of colored lines for sets of black holes with fixed values of $\phi_H$ (in this case $\phi_c=\phi_H$). The colour gradient indicates the size of the horizon radius $r_H$.
}
\label{plot_modes}
\end{figure}

In fact it is possible to obtain unstable radial modes for all the solutions we have studied. In Fig. \ref{plot_modes} we exhibit the spectra of unstable radial modes for hairy black holes and scalarons as a function of the $\phi_c$ parameter. The unstable modes decreases from zero to a minimum value as this parameter is increased. From the minimum, the unstable mode increases again as $\phi_c$ approaches the value of $\phi_1$ (the numerical results indicate that the mode becomes zero exactly at this value, not before). The modes for hairy black holes and scalarons are also smoothly connected in the small horizon limit. Interestingly we find that the unstable mode of the hairy black holes decreases in magnitude with the size of the horizon. This indicates that black holes with large horizon size could be effectively stable (at least in some time scale). 

\section{Conclusion}\label{sec:con}

The gravity minimally coupled to a scalar potential allows the construction of hairy black holes by requiring the scalar potential to be non-strictly positive for violating the energy conditions \cite{Herdeiro:2015waa}. Ref. \cite{Corichi:2005pa} employing such scalar potential which has a global minimum $\phi_1$, a local minimum $a$ and a local maximum $\phi_0$ to construct the spherically symmetric and asymptotically flat hairy black holes and mainly study their empirical mass formulae. The asymptotically flatness condition of black holes is guaranteed by fixing the local minimum of potential to be zero. This potential has been widely applied to study the quantum tunneling process from the false vacuum $a$ to the true vacuum $\phi_1$ in the cosmology.

In our paper, we perform a comprehensive study on the properties of black holes by solving the Einstein-matter field equations numerically. We fix the global minimum (true vacuum) and vary the value of scalar field at the horizon $\phi_H$ to generate the hairy black holes solutions. Thus, a branch of hairy black holes with fixed horizon size emerge from the Schwarzshild black holes. For small $\phi_1$, when $\phi_H$ increases from zero and approaches to $\phi_1$, the scaled area of horizon decreases from unity to zero and the scaled Hawking temperature increases very sharply from unity. %
In the limit $\phi_H=\phi_1$, the scalar field sits exactly at the true vacuum $\phi_1$ and no tunneling occurs, thus hairy black holes do not exist anymore in that limit. For large $\phi_1$, the scaled area of horizon also decreases from unity to zero but the scaled Hawking temperature increases to a finite value from unity when $\phi_H$ increases to a value which is still less than $\phi_1$. In that situation, we are unable to generate the solutions for $\phi_H$ beyond that value because the numerical code stop working. 

Analogous to the hairy black holes, a branch of globally regular particle-like solution which is known as scalaron emerges from the Minkowski spacetime by varying the scalar field at the origin $\phi_c$. The scalaron also behaves analogously with the hairy black holes in the limit $\phi_c = \phi_1$ where the mass increases very sharply. Similarly, we are also unable to generate the scalaron for large $\phi_1$. In addition, the hairy black holes are reduced to the scalaron in the small horizon limit. 

The profiles of both compact objects behave similarly, they have almost the constant functions in the bulk, which corresponding to the true vacuum $\phi_1$, then have a sharp boundary where the functions rapidly change to another set of constant functions which corresponding to the false vacuum $(a=0)$ lies at the infinity.  

We also investigate the linear stability of hairy black holes by performing the radial perturbation on the metric and scalar field. Hence, we obtain a master equation which is Schr\"odinger-like. We numerically solve the master equation to compute the spectra of radial modes. Generically both hairy black holes and scalaron are unstable against the radial perturbation. Both spectra decreases from zero to a minimum value and then increases to zero as the value of scalar field increases from zero and then approaches $\phi_1$. Moreover, the unstable modes of hairy black holes are connected with scalaron in the small horizon limit. We also find that the hairy black holes with larger horizon size are more relatively stable against the perturbation. %

There are many possible directions that can be derived from this paper. First, we can study the properties of non-asymptotically flat hairy black holes by not fixing the local minimum of potential as zero. We also can repeat the same approach to study the properties of hairy black holes by considering the non-minimally coupled scalar field with the Ricci scalar \cite{Nucamendi:1995ex}. Others are we can consider to study the properties of charged hairy black holes for this model and their linear stability. Since the astrophysical black holes are rotating, then it is interesting to study the difference between properties of rotating hairy black holes with Kerr black hole.

\section*{Acknowledgement}
 D.Y and XYC are supported by the National Research Foundation of Korea (Grant No.: 2021R1C1C1008622, 2021R1A4A5031460). XYC is grateful on the hospitality from the organizer at APCTP in Pohang to present this work in the workshop String theory, Gravity and Cosmology (SGC2022). JLBS gratefully acknowledges support from MICINN project PID2021-125617NB-I00, Santander-UCM project PR44/21‐29910, DFG Research Training Group 1620 \textit{Models of Gravity} and FCT project PTDC/FIS-AST/3041/2020. We are grateful to have a useful discussion with Jutta Kunz.

\end{document}